\documentclass[%
 aip,
 amsmath,amssymb,
preprint,%
]{revtex4-2}

\usepackage{graphicx}
\usepackage{dcolumn}

\usepackage{multirow}
\usepackage[dvipsnames]{xcolor}
\usepackage{makecell}
\usepackage{hyperref}

\usepackage[utf8]{inputenc}
\usepackage[T1]{fontenc}
\usepackage{mathptmx}
\usepackage{etoolbox}
\usepackage{xcolor}
\usepackage{setspace}
\usepackage[ruled, lined, linesnumbered, commentsnumbered, longend]{algorithm2e}

\makeatletter
\def\@email#1#2{%
 \endgroup
 \patchcmd{\titleblock@produce}
  {\frontmatter@RRAPformat}
  {\frontmatter@RRAPformat{\produce@RRAP{*#1\href{mailto:#2}{#2}}}\frontmatter@RRAPformat}
  {}{}
}%
\makeatother
\begin{document}

\title{Model adaptive phase space reconstruction}
\author{Jayesh M. Dhadphale}
\email{jayeshmdhadphale@gmail.com}
\affiliation{Department of Aerospace Engineering, Indian Institute of Technology Madras,
Chennai, Tamil Nadu 600036, India}
\author{K. Hauke Kraemer}%
\affiliation{Potsdam Institute for Climate Impact Research, Member of the Leibniz Association, 14473 Potsdam, Germany}%

\author{Maximilian Gelbrecht}%
\affiliation{Potsdam Institute for Climate Impact Research, Member of the Leibniz Association, 14473 Potsdam, Germany}
\affiliation{School of Engineering \& Design, Technical University of Munich, Munich, Germany}%

\author{Jürgen Kurths}%
\affiliation{Potsdam Institute for Climate Impact Research, Member of the Leibniz Association, 14473 Potsdam, Germany}%

\author{Norbert Marwan} 
\affiliation{Potsdam Institute for Climate Impact Research, Member of the Leibniz Association, 14473 Potsdam, Germany}
\affiliation{Institute of Physics and Astronomy, University of Potsdam, Germany}%
\author{R. I. Sujith}
\email{sujith@iitm.ac.in}
\affiliation{Department of Aerospace Engineering, Indian Institute of Technology Madras,
Chennai, Tamil Nadu 600036, India}

\date{\today}

\begin{abstract}
    Phase space reconstruction methods allow for the analysis of low-dimensional data with methods from dynamical system theory, but their application to prediction models, like those from machine learning, is limited.  
    Therefore, we present here a model adaptive phase space reconstruction (MAPSR) method that unifies the process of phase space reconstruction with the modeling of the dynamical system. 
    MAPSR is a differentiable phase space reconstruction (PSR) method that enables the use of machine learning (ML) methods and is based on the idea of time delay embedding. 
	For achieving differentiable, continuous, real-valued delays, which can be optimized using gradient descent, the discrete time signal is converted to a continuous time signal.
    The delay vector, which stores all potential embedding delays and the trainable parameters of the model are simultaneously updated to achieve an optimal time delay embedding for the observed system.
    MAPSR does not rely on any threshold or statistical criterion for determining the dimension and the set of delay values for the embedding process. 
    The quality of the reconstruction is evaluated by the prediction loss. 
    We apply the proposed approach to uni- and multivariate time series stemming from regular and chaotic dynamical systems and a turbulent combustor to test the generalizability of the method and compare our results with established phase space reconstruction methods\cite{paper:rc_pathak_2018, book:kantz_schreiber_2003, paper:parlitz_2000, paper:b_lim_2021}.
    We find that for the Lorenz system, the model trained with the MAPSR method is able to predict chaotic time series for nearly 7 to 8 Lyapunov time scales which is found to be much better compared to other PSR methods (AMI-FNN and PECUZAL methods). 
    For the univariate time series from the turbulent combustor, the long-term prediction error of the model trained using the MAPSR method stays in between that of AMI-FNN and PECUZAL methods for the regime of chaos, and for the regime of intermittency, the MAPSR method outperforms the AMI-FNN and PECUZAL methods.
\end{abstract}

\maketitle

\begin{quotation}

\end{quotation}

\section{Introduction}
\label{sec:introduction}
The evolution of deterministic dynamical systems is governed by a set of rules \cite{book:chaos_1997}. 
The quest to discover these rules has led to various discoveries in science. 
These rules can be identified by deriving the mathematical expressions starting with the first principles \cite{book:chaos_1997}. 
This approach is tedious for dynamical systems with large degrees of freedom, and obtaining predictions is computationally expensive. 
However, dynamical systems with such large degrees of freedom often exhibit dynamics in a much smaller subset of the entire state space \cite{book:chaos_1997, paper:mullin_1991, paper:mullin_1995}.

 A vector in the system's state space, the state vector, defines the dynamical state of the system. 
 In practice, the inaccessibility of a dynamical system often limits the number of measured state variables and, therefore, results in an incomplete state vector.
 In those cases it is nevertheless possible to reconstruct the attractor of the unknown state space according to the embedding theorems of \citet{paper:whitney_1936}, \citet{paper:mane_1981}, and \citet{paper:takens_1981} using different techniques such as derivative coordinates \cite{paper:BROOMHEAD1986217,  paper:MANN20112999}, Legendre coordinates \cite{paper:GIBSON19921}, and delay coordinates \cite{paper:packard_1980}. 

The attractor reconstructed from the measured time series data has a similar topology as that of the measured dynamical system \cite{paper:CASDAGLI199152}, i.e. is diffeomorphic to it. 
The properties, such as the Lyapunov exponent, eigenvalues of fixed points or the fractal dimension, can be preserved under the phase space reconstruction \cite{paper:CASDAGLI199152} (PSR). 
PSR attempts to create an attractor with a sufficient embedding dimension to avoid the intersection of the trajectories and guarantee a diffeomorphic mapping. 
According to Taken's theorem \cite{paper:takens_1981}, the embedding can be achieved if the reconstructed phase space has a dimension ($D$) greater than twice that of the box-counting dimension ($D_B$) of the actual, unknown, dynamical system; i.e., $D>2D_B$. 

However, Taken's theorem\cite{paper:takens_1981} assumes a clean time series of length $N\to \infty$ and a sampling time $\Delta t \to 0$ to guarantee a diffeomorphic mapping for any delay value (except for some pathological, periodic cases).
For real-world time series, time delay reconstruction methods try to balance too small delay values, which lead to \textit{redundancy}, and too large delay values, which lead to \textit{irrelevance} of coordinates \cite{paper:CASDAGLI199152, paper:uzal_2011, paper:ROSENSTEIN199482, paper:armin_2018}. 
Since noise is present in real-world time series data, the choice of appropriate delay values is important to avoid amplification of the noise and to keep the complexity of the attractor within limits \cite{paper:CASDAGLI199152, paper:uzal_2011}.  

The time series $\vec{s}(t)=[s_i(t);\; i=1,\ldots, m]$ measured from the dynamical system can be univariate ($m=1$) or multivariate ($m>1$). For univariate time series, the delay coordinates with a dimension of $d$ can be represented as $\vec{x}(t)=[s_1(t+\tau_j);\; j=1,\ldots, d]$. 
The set of delays can be uniformly spaced, i.e., $\Delta\tau=\tau_{i+1}-\tau_i = \text{const.} \forall i$ (known as uniform time delay embedding (UTDE)) or non-uniformly spaced (NUTDE). 
In UTDE delays and embedding dimension are usually estimated, e.g., using average mutual information and false nearest neighbor crtieria \cite{paper:kennel_1992, paper:kennel_2002, paper:CAO199743, paper:hegger_1999, paper:Krakovsk2015UseOF}. 
For NUTDE, recent work by \citet{paper:Kraemer_2021} proposes a method (\emph{PECUZAL}) that unifies the continuity statistic of \citet{paper:pecora_1995} and \citet{paper:pecora_2007} quantifying functional dependence, with the $L$-statistic of \citet{paper:uzal_2011}, which quantifies the noise amplification. The former can be seen as a delay estimator and the latter as a dimension estimator given those estimated delays (we refer the interested reader to \citet{paper:Kraemer2022}).
NUTDE and UTDE techniques usually optimize an objective function that quantifies the goodness of the reconstruction, such as the $L$- or false nearest neighbor-statistic, \citep{paper:Kraemer2022} proposed to solve this optimization with a decision tree search.
\citet{paper:tan_2023} propose a method based on persistent homology intending to get delay values for NUTDE, which are independently selected and have dynamical explainability. In addition, these authors provide a brief overview of the embedding techniques.

\citet{paper:brunton_2017} proposed to model the dynamics of the chaotic system as an intermittently forced linear system that combines delay embedding and Koopman theory \cite{paper:koopman_2013}.
The intermittent forcing is required when the dynamics is strongly nonlinear and needs to be determined from the time series. 
\citet{paper:bakarji2022discovering} use UTDE to get an interpretable closed-form expression of the dynamical system. 
An encoder maps the reconstructed attractor to low dimensional space, where a closed-form model is obtained for this encoded attractor using the SINDy method \cite{paper:brunton_2016}. 

Ultimately, one goal of PSR approaches is to predict the system in question, given its incomplete observations. 
Prediction based on PSR can, e.g., use models that extrapolate based on neighborhoods in the reconstructed phase space. 
Several approaches have been made that differ in the exact way a local neighborhood-based model is built \cite{Farmer1987,Casdagli1989,Sugihara1990,Kantz2004,Isensee2020}. 
Alternatively, \citet{paper:jayesh_2022} used a delay embedding technique along with a neural ODE approach in order to yield a suitable model for a thermoacoustic system \cite{book:sujith_2021}. 
Neural ODEs \cite{paper:Chen_NODE_2018_7892} are a natural candidate for data-driven modeling of dynamical systems. 
They integrate artificial neural networks (ANNs) into the right-hand side of differential equations. 
As ANNs are universal function approximators, a neural ODE is trivially a universal dynamical system approximator. 
What kind of PSR is optimal in this case? \citet{paper:Kraemer2022} were optimizing the PSR for predictions with a decision tree search. 
However, this approach is costly. 
In this article, we therefore propose an alternative: a differentiable variant of a time-delay embedding that makes use of ANNs and optimizes for the prediction loss.
The method simultaneously updates the time delays $\vec{\tau}$ and the trainable parameters (excluding hyperparameters) of the mathematical model for the system to minimize the assumed objective function. 
Hence, we name the method as model adaptive phase space reconstruction (MAPSR). 
We showcase this approach with neural ODEs, but the MAPSR method fits perfectly in machine learning frameworks and can readily be used with other data-driven models. 

Three other frameworks can be thought of as related to our proposal. 
The reservoir computing (RC) framework can also be used to model the dynamical system from available low-dimensional data \cite{paper:rc_pathak_2018, paper:rc_LUKOSEVICIUS2009127}. 
For resemblence of RC with the delay embedding, we refer to \citet{paper:rc_PhysRevResearch.5.L022041}. 
The latent ODE framework \cite{paper:latent_ode_2019} is a method for sequence-to-sequence learning that tries to learn an ODE in an adaptively learned latent space. Augmented neural ODEs \cite{paper:dupont_2019} add unobserved, latent dimensions to an otherwise unmodified neural ODE. Compared to these methods, our approach, MAPSR, also achieves an interpretable PSR that we can investigate further. 

The remainder of this paper is organized as follows: 
The proposed methodology of the MAPSR method is described in Sec.~\ref{sec:MAPSR_description}.
Sec.~\ref{sec:results_and_discussion} presents the results and discusses how the MAPSR performs on standard dynamical systems and on time series obtained from real-world dynamical systems. 
The key features and limitations of the method are summarized in Sec.~\ref{sec:conclusion}. The brief algorithm of the MAPSR is presented in Appendix \ref{app:MAPSR_algo}.

\section{Description of the method}\label{sec:MAPSR_description}

The first step in nonlinear time series analysis is often the PSR from available time series data \cite{book:kantz_schreiber_2003}. 
The PSR is conventionally performed by targeting the independence of the selected coordinates by optimizing an objective function, which reflects such an independence \cite{paper:kennel_1992, paper:kennel_2002, paper:CAO199743, paper:hegger_1999, paper:Krakovsk2015UseOF, paper:pecora_1995, paper:pecora_2007}. At the same time the obtained reconstructed trajectory is not optimized for a specific application or analysis.
\citet{paper:Kraemer2022} have discussed this issue, provided a modular way to choose the statistic and objective function for the delay selection according to the research question, and used the MCDTS (Monte Carlo Decision Tree Search) method to obtain global minima.  
The objective of the current work is to make the parameters of the phase space reconstruction differentiable to avoid a combinatorial selection of delay values. 
This step allows us to use a common optimization framework to determine the optimal parameters for the phase space and trainable parameters of the assumed mathematical model of the dynamical system.
Specifically, we intend to create an initial delay vector $\vec{\tau}$ of a certain initial dimension $D_{init}$, which gives us an initial trajectory in our reconstruction phase space. 
We will use this trajectory for training a mathematical model specifically an ANN, which can approximate the underlying ordinary differential equation from which the trajectory can be obtained by integration. 
The prediction error for a certain prediction horizon will serve as the loss function.
The model will, therefore, depend on $\vec{\tau}$ and optimizes this vector along with its own parameters via a gradient descent method. 
We allow the delay vector to reduce in size, i.e., reduction of $D_{init}$, during training. 

Conventional methods for PSR attempt to find the delays that are multiples of the sampling time $\Delta t$ \cite{paper:kennel_1992, paper:kennel_2002, paper:CAO199743, paper:hegger_1999, paper:Krakovsk2015UseOF}. 
Here, we remove this restriction by converting the discrete-time measurements into continuous-time variables using interpolation, which allows delays to take continuous values. 
The advantage of this conversion is that a continuous variable is now piecewise differentiable. 
Suppose the measured time series $\vec{s}(t)$ is a multivariate vector time series with $m$ components or variables, i.e., $\vec{s}(t)=[s_i(t)| i=1,\ldots, m]$. 
We define the delay vector 
\begin{equation}
    \vec{\tau}=[\tau_{1,1},\tau_{1,2},\ldots,\tau_{1,d_1},\tau_{2,1},\ldots,\tau_{2,d_2},\ldots,\tau_{m,d_m}],
    \label{eq:delay_vector}
\end{equation}
where $\tau_{i,j}$ is the $j^{th}$ delay associated with the $i^{th}$ measured variable. 
Here, for the $i^{th}$ measured variable, there are $d_i$ delay values; i.e., $\vec{\tau}$ has $D=\sum_i d_i$ components, and $D$ corresponds to the dimension of the phase space. 
The vector $\vec{\tau}$ is initialized such that the first delay value associated with all the time series is set to zero; i.e., $\tau_{i,1}=0, \forall i$. 
The remaining components of the $\vec{\tau}$ are initialized assuming UTDE for individual time series. 
For example the time delay ($\tau_{i,AMI}$) and dimension ($d_{i,AMI}$) obtained from the first minimum of the auto-mutual information $AMI(\tau)$ together with a dimension estimator like the FNN-statistic, provides estimates for the order of the common difference $\Delta \tau_i$ and the initial dimension ($d_{i}$) for each time series. 
This means that while initializing $\vec{\tau}$ the common difference between successive delay values, $\Delta \tau_i = \tau_{i,j+1}- \tau_{i,j}$, for the $i^{th}$ time series is maintained constant such that $\Delta \tau_i \sim O(\Delta \tau_{i,AMI})$. 
Note that UTDE is only used to initialize the $\vec{\tau}$ vector before training.
$\vec{\tau}$ is initialized with $(d_i)_{init}=d_{init}$, and the common difference $\Delta \tau$ is the same for all time series. 
For simplicity, we will refer to $d_{init}$ in the subsequent plots; i.e., the initial dimension of the delay vector is $D_{init}=\sum_{i=1}^{m} (d_i)_{init}=d_{init}m$, where $m$ is the number of time series. 
In practice, we train the model for different initial dimensions $d_{init}=1 \dots d_{\max}$ and eventually the model with the initial dimension $d_{init}$ with minimum loss after the training, i.e., $\mathcal{L}_{min}(d_{init})$ gives the optimal set of delays.
$d_{\max}$ can be set to an arbitrary, yet large enough value or can be of the order of the dimension estimated using the AMI-FNN method for an educated guess.

Further below, we describe how the delay values may merge during training (c.f.Sec.~\ref{subsubsec:merging}). Thus, the final number of delay values for each time series $(d_i)_{final}$ and $D_{final}$ might be less than the $D_{init}$. 
The initially chosen dimension of $\vec{\tau}$ must be chosen sufficiently large. 
When set too small, we do not expect the model to perform well, which will be reflected in the training loss $\mathcal{L}(d_{init})$ not achieving a minimum with respect to other training based on larger $D_{init}$.
Our expectation is that for a sufficiently high initial dimension $D_{init}$ the training loss $\mathcal{L}(d_{init})$ is minimal.

\subsection{Training the model}\label{subsec:training}

With MAPSR, we now allow $\tau_{i,j}$ to be a non-integer multiple of the sampling time $\Delta t$. 
Then, the time series $\vec{s}(t)$ is interpolated to get the vector of delay coordinates with these non-integer delays. 
At time $t=n\Delta t$ where $n$ is a non-negative integer, the vector of delay coordinates is $\vec{x}(n\Delta t, \vec{\tau})=[s_1(\tau_{1,1}+n\Delta t),\ldots,s_1(\tau_{1,d_1}+n\Delta t), s_2(\tau_{2,1}+n\Delta t),\ldots,s_2(\tau_{2,d_2}+n\Delta t),\ldots,s_m(\tau_{m,d_m}+n\Delta t)]$. 
In this paper, we have used linear interpolation to compute $s_i(t+\tau_j)$, but other interpolation techniques can be easily incorporated as well. 

For example, $s_i(t+\tau_{i,j})$ can be computed using linear interpolation as,
\begin{equation}
    s_i(t+\tau_{i,j})= (1-\beta)s_i(h\Delta t) + \beta s_i((h+1)\Delta t),
    \label{eq:interpolation}
\end{equation}

\noindent where $\beta=(t+\tau_{i,j}-h\Delta t)/\Delta t$ and time instance $(t+\tau_{i,j})$ lies between $h^{th}$ and $(h+1)^{th}$ sampling instances, i.e., $h\Delta t\leq (t+ \tau_{i,j}) <(h+1)\Delta t$. 
The interpolation gives the time series of $\vec{x}(n\Delta t, \vec{\tau})$ with the sampling time $\Delta t$ being identical to the sampling time of the original time series. 

After the first step of defining the delay vector and obtaining the delay coordinates, the next step in MAPSR is modelling the dynamics. 
The modelling aims to determine the function $\vec{f}$ such that $\dot{\vec{x}}(t)=\vec{f}(\vec{x}(t, \vec{\tau}),W)$, where $\dot{\vec{x}}$ is the time derivative of the reconstructed state vector $\vec{x}$, and $W$ is the set of parameters that governs the dynamical behavior of the model. 
The model can be linear or nonlinear where $\vec{f}$ is differentiable with respect to $W$ and $\vec{\tau}$. The functional form of $\vec{f}$ is usually unknown, but can be approximated by universal function approximators such as ANNs \cite{paper:Kur_1989_Universal_Approx}. ANNs that approximate the right-hand side of an ODE are known as neural ODEs \cite{paper:Chen_NODE_2018_7892}. 
Thus, the differential equation for the trajectories yielded from the given time series time-delay embedded with the delay vector $\vec{\tau}$, Eq.~\eqref{eq:delay_vector}, can be approximated with a neural ODE. 
For the rest of the paper, $\vec{f}$ is approximated with neural ODE, but the proposed method is not limited to neural ODE and allows for other methods as well.

\begin{figure*}[ht!]
    \centering\linespread{2}
    \includegraphics[width=0.9\textwidth]{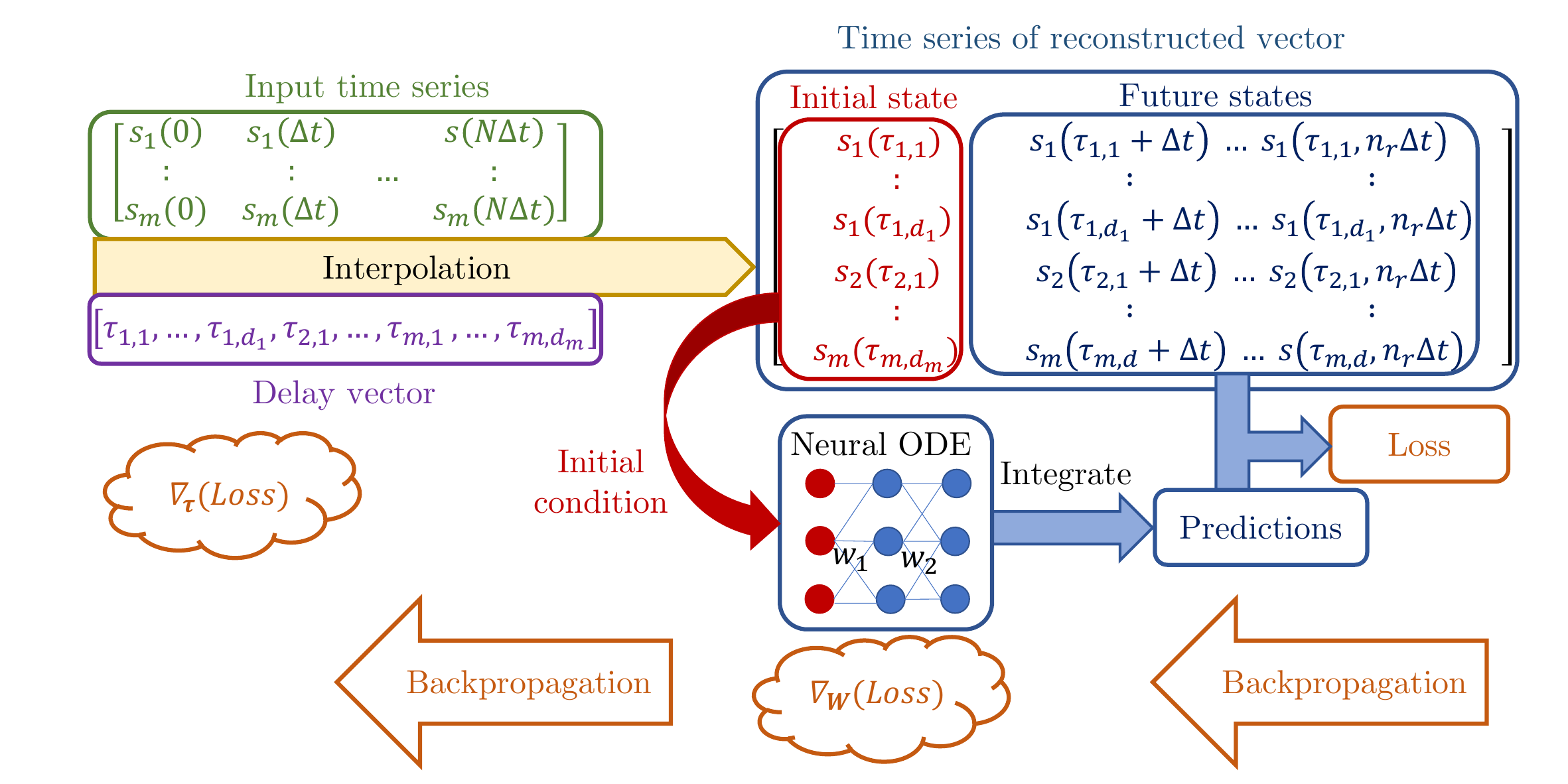}
     \caption{Flow chart of the model adaptive phase space reconstruction during training. 
     Arrows from left to right (forward pass) show the steps involved in the computation of the loss function. 
     During the backpropagation, the gradient of the loss is computed with respect to $W$ and $\vec{\tau}$.}
     \label{fig:mapsr_overview}
\end{figure*}

The performance of the trained model and the goodness of the reconstructed phase space can be tested by time integrating the model and quantifying the prediction loss. 
The time series of the delay coordinates obtained after interpolation gives the initial state of the system as $\vec{x}(t_0)$ (Fig.~\ref{fig:mapsr_overview}). 
The future state of the system after $r$ time steps, i.e., at time $t_0 + r\Delta t$ is predicted as
\begin{equation}
    \vec{x}_{r,pred}=\vec{x}_{pred}(t_0+r\Delta t, \vec{\tau}) =\vec{x}(t_0)+ \int_{t_0}^{t_0+r\Delta t}f(\vec{x}(t, \vec{\tau}),W) dt.
    \label{eq:mapsr_pred}
\end{equation}
Let the prediction horizon be $T_R=R\Delta t$, where $R=\max(r)$, is fixed to be of the order of a characteristic time scale of the system, such as the Lyapunov time scale or the period in the case of limit cycle oscillations (LCO). 

Suppose $K$ points are randomly chosen from the reconstructed trajectory, and the $k^{th}$ selected point corresponds to the time instance $t_k$. 
The state of the dynamical system at each of these $K$ points is treated as the initial condition, and future states are predicted $R$ time steps ahead. 
The variable $K$ is commonly referred to as batch size. 
The loss in the prediction is computed by averaging the prediction error as
\begin{equation}
    Loss = \mathcal{L}(R, d_{init}) = \bigg<\sum_{r=1}^{R}\left\|\vec{x}_{r,true}-\vec{x}_{r,pred}\right\|_p^p\bigg>_k
    \label{eq:mapsr_loss}
\end{equation}
and depends on the chosen prediction horizon $T_R=R\Delta t$. Here, $\|\cdot\|_p$ is the $p^{th}$ norm and $\langle\,\cdot\,\rangle_k$ indicates the average over the batch. 
The loss function quantifies the prediction error of the model by comparing the predicted states $\vec{x}_{r,pred}$ using the model with the future states obtained with interpolation of the measured time series $\vec{x}_{r,true}$. 

The above description shows the steps in the forward pass, i.e., how the delay vector is initialized, the model is used to perform prediction, and how the loss is computed.
The MAPSR method aims to optimize $\mathcal{L}(R, d_{init})$ with respect to $\vec{\tau}$ and $W$.
This optimization needs the gradient of the loss function with respect to $\vec{\tau}$ and $W$ as $\nabla_{\vec{\tau}}(\mathcal{L})$ and $\nabla_{W}(\mathcal{L})$, respectively.
Backpropagation is one of the methods to compute the derivative of the loss function with respect to the differentiable parameters by back-tracing all the operations starting from the loss \cite{review:baydin2018automatic}.
The introduction of the interpolation method to compute delay coordinates allows the backpropagation algorithm to calculate $\nabla_{\vec{\tau}}(\mathcal{L})$ and also allows to apply this method to non-equidistantly sampled time series or time series with (small) gaps, which is the crucial idea we introduce in this paper.
The loss function is minimized by iteratively updating the $\vec{\tau}$ and $W$ using the RMSprop algorithm \cite{notes:tieleman2012lecture}. 
The minimum component of the delay vector is subtracted from itself to maintain the minimum delay value as zero and all other delay values as non-negative. 
Thus $\vec{\tau}$ is redefined as,

\begin{equation}
        \vec{\tau} := \vec{\tau} - \min(\vec{\tau}).
    \label{eq:mapsr_tau_update}
\end{equation}

\noindent after each optimization step.

\subsection{Dimension reduction while training}\label{subsubsec:merging}

During the iterative update of the delay vector, if two delay values associated with the $i^{th}$ time series, say $\tau_{i,j}$ and $\tau_{i,k}$, are closer than a certain threshold $\tau_{th}$, i.e., $|\tau_{i,j}-\tau_{i,k}|<\tau_{th}$, one of the delay values is removed from the $\vec{\tau}$. 
Removing the delay values is equivalent to removing the associated delay coordinates from the PSR, thus, decreasing the dimension $D$ of the reconstructed state space. 
We term this as delay \emph{merging}. 
Very close delay values are associated with \emph{redundant} delay coordinates, which do not convey any additional information about the state of the system and can, therefore, be removed. 
After merging the delays, the dimension of the embedding vector $\vec{x}$ reduces. 
Suppose the weight matrix associated with the first layer of the neural ODE is $W_1$ which maps the input vector $\vec{x}$ as $\vec{z}_1 = {W_1}^T\vec{x}.$
If $x_i$ and $x_j$ merge, to avoid restarting the training from scratch, the $i^{th}$ row in the weight matrix $W_1$ is replaced by the addition of the $i^{th}$ and the $j^{th}$ rows and further the $j^{th}$ row is removed from the matrix. 
The $j^{th}$ column from the weight matrix associated with the last layer of neural ODE is also removed to reduce the dimension of the output vector. 
This modification of the weight matrices safely reduces the number of nodes in the input and output layer, i.e., reduces the embedding dimension without affecting the neural ODE model.
During training, the delay values may merge, and the final number of delay values $(d_i)_{final}$ might be less than the $d_{init}$. 
As mentioned earlier, the minimum loss that can be achieved during training does not only depend on the prediction horizon $R$, but also on the initial dimension of $\vec{\tau}$. 
The optimal delay embedding is selected based on the minimum of the $\mathcal{L}(d_{init})$,
\begin{equation}
    \mathcal{L_{\min}}(d_{init}^{(opt)}) = \min_{d_{init}\in[1,d_{\max}]} \mathcal{L}(d_{init})
\end{equation}
The optimal initial dimension $d_{init}^{(opt)}$ for which loss is minimum gives the optimal delay embedding,
\begin{equation}
    \vec{\tau}^{(opt)}_{init}=[\tau_{1,1},\tau_{1,2},\ldots,\tau_{1,d_{init}^{(opt)}},\tau_{2,1},\ldots,\tau_{2,d_{init}^{(opt)}},\ldots,\tau_{m,d_{init}^{(opt)}}]
    \label{eq:init_delay_opt}
\end{equation}
and the delay vector after training can be written as
\begin{equation}
    \vec{\tau}_{final}^{(opt)}=[\tau_{1,1},\tau_{1,2},\ldots,\tau_{1,d_{1,final}^{(opt)}},\tau_{2,1},\ldots,\tau_{2,d_{2,final}^{(opt)}},\ldots,\tau_{m,d_{m,final}^{(opt)}}].
    \label{eq:final_delay_opt}
\end{equation}

From Eq.~(\ref{eq:init_delay_opt}) we can see that initially for all time series $(d_i)_{init}=d_{init}^{(opt)}$ and from Eq.~(\ref{eq:final_delay_opt}) we can see that after training $(d_i)_{final}=d_{i,final}^{(opt)}$. 
Thus, if there are $m$ time series, then due to merging of the delays $\sum_i d_{i, final}^{(opt)}\leq m d_{init}^{(opt)}$.
The Algorithm \ref{alg:MAPSR} gives the step-by-step description of the MAPSR method.

The expected behavior of the loss function is that it decreases with increasing the initial dimension $d_{init}$ until it is sufficiently large so that trajectories do not intersect anymore.
Increasing $d_{init}$ further is not expected to result in any further decrease of the loss function. 
Contrastingly, our investigation reveals that the loss function increases after the optimal dimension.
For the $D$ dimensional PSR all the $D$ components of the $(\vec{x}_{r,true}-\vec{x}_{r,pred})$ contribute to the loss function Eq. \ref{eq:mapsr_loss}.
Below the optimal dimension, the trajectories cannot be sufficiently resolved, and the loss decreases initially with the initial dimension.
Once the trajectories are well resolved and can be captured by the model, the loss function attains minima.
There is a discrepancy between the trajectories predicted by the model and true trajectories due to imperfect modeling or noise in the data. 
Adding the new delay coordinate after the optimal dimension might just increase this discrepancy in the loss function and cause it to increase.
This contribution to the loss function keeps increasing upon adding new delay coordinates, and therefore the loss keeps growing after the optimal dimension. 
We discuss this in more detail in Sec.~\ref{subsec:analysis}. 

\section{Results and discussion} \label{sec:results_and_discussion}

We test the MAPSR method on the time series from selected dynamical systems, such as univariate time series from a harmonic oscillator and the Lorenz system, and also on univariate and multivariate time series acquired from a real system, here a turbulent combustor.
The following subsections present the results obtained using the MAPSR method for univariate and multivariate time series data.

\subsection{Application of the MAPSR method to univariate time series from the harmonic oscillator}
A univariate time series for the harmonic oscillator is generated by the expression, $s_1(t_n) = \sin(2\pi t_n)+ \epsilon$, where $\epsilon\sim \mathcal{N}(0,\sigma^2)$ represents noise, i.e., $\epsilon$ is normally distributed with zero mean and variance $\sigma^2$. 
We use six noise levels, i.e., clean signal ($\sigma=0$) and noisy signals with $\sigma$ equal to 1\% to 5\% in steps of 1\% of the amplitude of $s_1$. 
The amplitude is computed as the mean of $|\max(s_1)|$ and $|\min(s_1)|$. The time series is normalized to have zero mean and a maximum absolute value of one. 
The time series is evenly sampled with the sampling period $\Delta t=0.01$ s.

\begin{figure*}[ht!]
    \centering\linespread{2}
    \includegraphics[width=\textwidth]{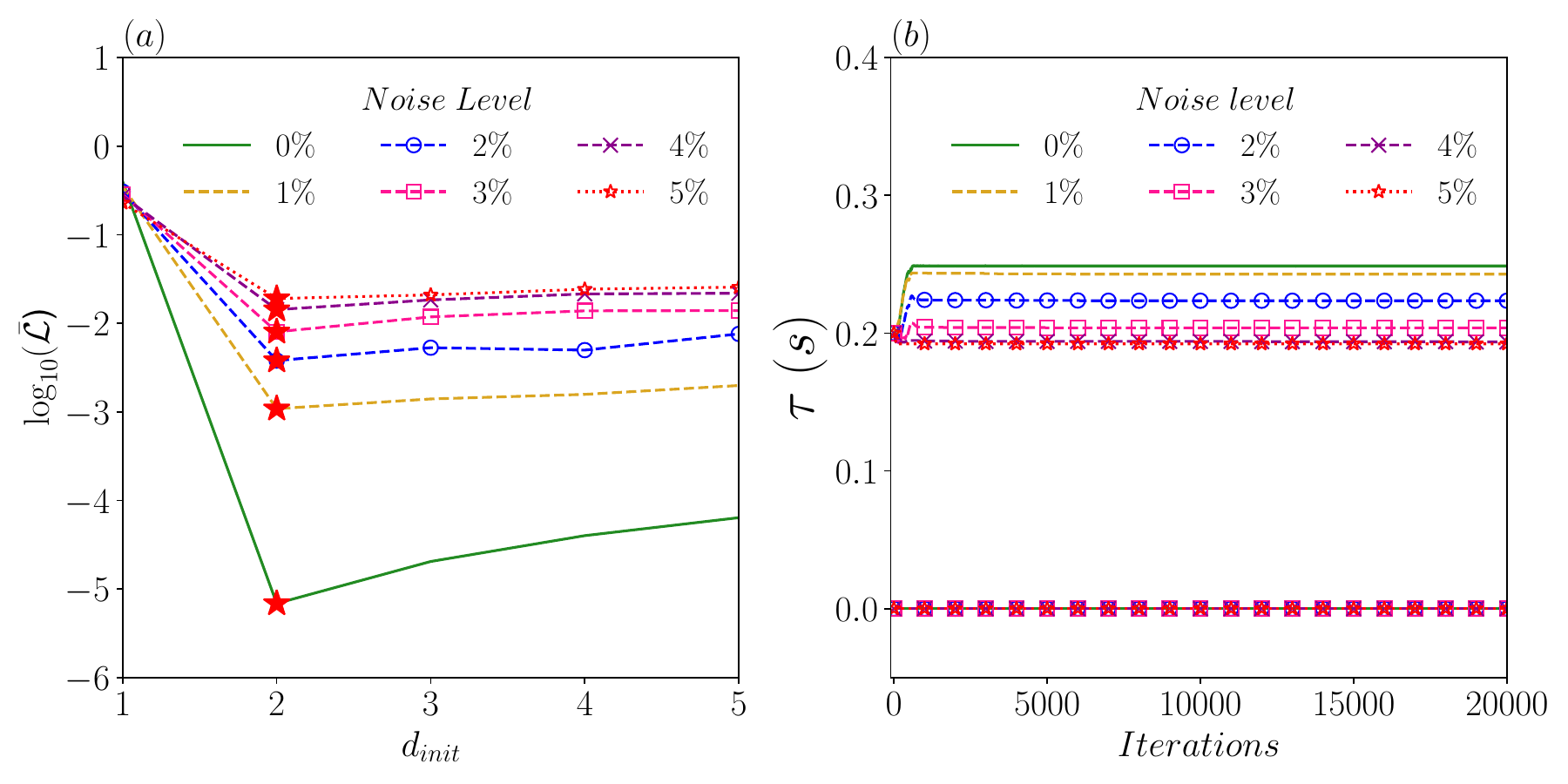}
    \caption{Application of the MAPSR method to time series of a harmonic oscillator with time period $T=1$. 
    (a) Variation of the prediction loss in log scale $\log_{10}(\mathcal{\bar{L}})$ with the initial dimension of the delay vector. 
    The MAPSR is applied to time series with six noise levels, i.e., $\sigma= 0\%$-$5\%$ of the amplitude of the signal. 
    The dimension at minima of $\log_{10}(\mathcal{\bar{L}})$ is selected as the embedding dimension ({\color{red}$\star$}). 
    (b) The evolution of the selected delay vector for the optimal embedding dimension. 
    Here, the $d=2$ is the optimum dimension for all noise levels. 
    The delay vector was initialized as $[0,0.20]$ s for all the noise levels and updated during training. 
    For low noise level, the delay vector approaches $[0,0.25]$ s, which corresponds to linearly independent coordinates.}
    \label{fig:harmonic_1}
\end{figure*}

\begin{table*}[ht]
    \centering\linespread{2}\renewcommand{\arraystretch}{1.5}
\begin{tabular}{|p{3cm}||p{2cm}|c|p{7cm}|}
\hline
    Case & Method & Dimension &  Delay (s)\\
    \hline\hline
    \multirow[c]{3}{*}{Noise level: 0\%} & AMI-FNN & 7 & [0.0  , 0.25, 0.5 , 0.75, 1.0  , 1.25, 1.5 ]
 \\
 & MAPSR & 2 & [0.0    , 0.2488]
 \\
 & PECUZAL & 2 & [0.0  , 0.25]
 \\\hline
\multirow[c]{3}{*}{Noise level: 1\%} & AMI-FNN & 4 & [0.0  , 0.25, 0.5 , 0.75]
 \\
 & MAPSR & 2 & [0.0    , 0.2428]
 \\
 & PECUZAL & 2 & [0.0  , 0.25]
 \\\hline
\multirow[c]{3}{*}{Noise level: 2\%} & AMI-FNN & 4 & [0.0  , 0.25, 0.5 , 0.75]
 \\
 & MAPSR & 2 & [0.0    , 0.2235]
 \\
 & PECUZAL & 3 & [0.0  , 0.25, 0.23]
 \\\hline
\multirow[c]{3}{*}{Noise level: 3\%} & AMI-FNN & 4 & [0.0  , 0.25, 0.5 , 0.75]
 \\
 & MAPSR & 2 & [0.0    , 0.2038]
 \\
 & PECUZAL & 4 & [0.0  , 0.25, 0.23, 0.02]
 \\\hline
\multirow[c]{3}{*}{Noise level: 4\%} & AMI-FNN & 4 & [0.0  , 0.25, 0.5 , 0.75]
 \\
 & MAPSR & 2 & [0.0    , 0.1936]
 \\
 & PECUZAL & 2 & [0.0  , 0.25]
 \\\hline
\multirow[c]{3}{*}{Noise level: 5\%} & AMI-FNN & 4 & [0.0  , 0.25, 0.5 , 0.75]
 \\
 & MAPSR & 2 & [0.0    , 0.1922]
 \\
 & PECUZAL & 3 & [0.0  , 0.25, 0.23]\\
     \hline
\end{tabular}
\caption{Embedding dimension and delay values for the harmonic oscillator with different noise levels, estimated with different phase space reconstruction methods, i.e., AMI-FNN, MAPSR, and PECUZAL.}
    \label{tab:harmonic}
\end{table*}

\begin{figure*}[ht!]
    \centering\linespread{2}
    \includegraphics[width=\textwidth]{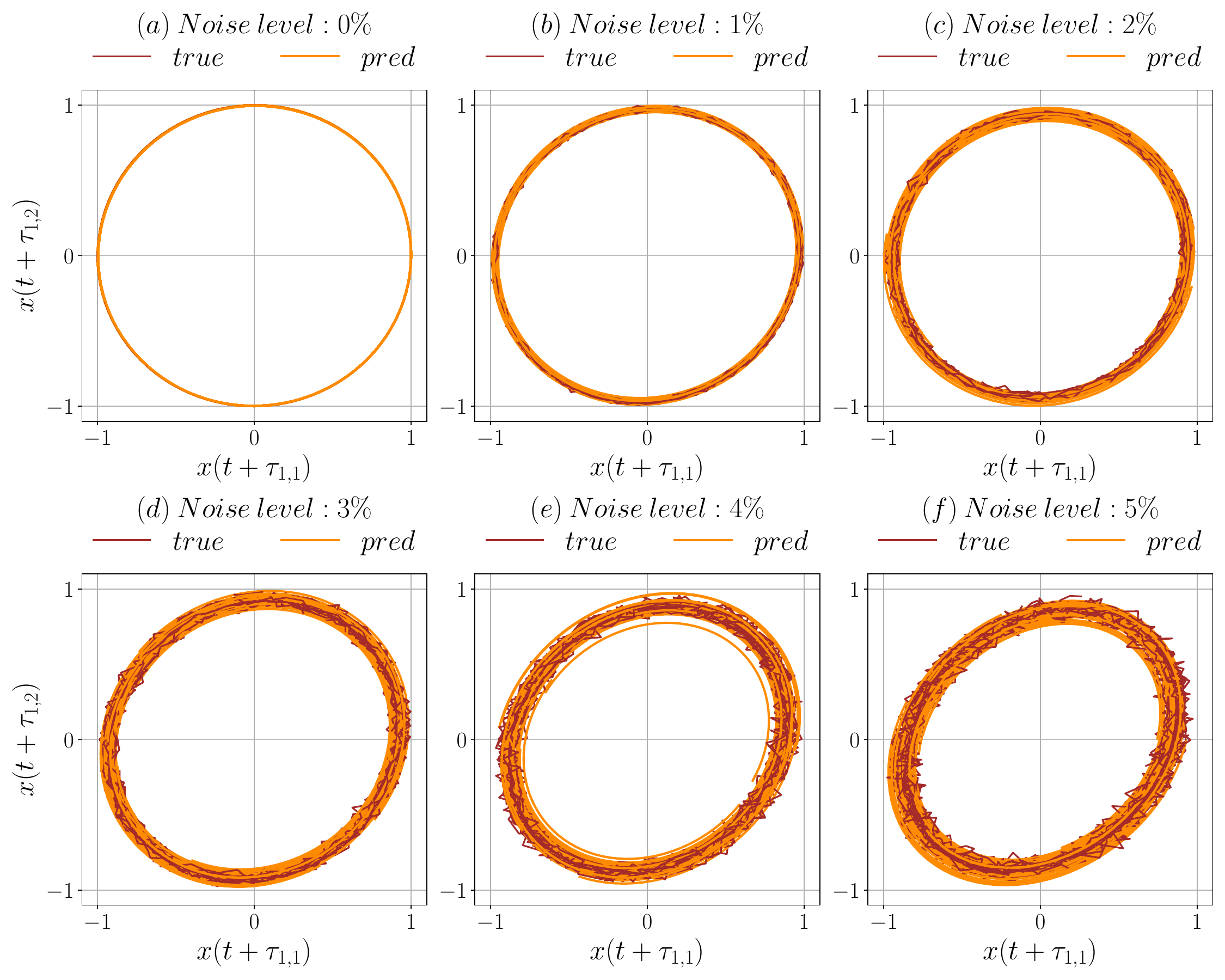}
    \caption{Comparison of the true and predicted trajectories using MAPSR method for six different noise levels 0\%-5\% for the harmonic oscillator.}
    \label{fig:harmonic_phase_portrait}
\end{figure*}

The dynamical system, $\dot{\vec{x}}=\vec{f}(\vec{x}, W)$ is known to be linear for harmonic oscillator; hence a neural ODE with direct linear mapping from the input to the output is used without any nonlinear activation function. 
Hence, $\dot{\vec{x}}={W_1}^T\vec{x}+\vec{b}_1$, here, $W_1$ is the weight matrix and $\vec{b}_1$ is the bias. 
For the $D$ dimensional phase space, the weight matrix $W_1$ has a dimension of $D\times D$. 
The weight matrix is randomly initialized before training.

For the time series of the harmonic oscillator considered in the current study, the AMI-FNN method predicts the optimal delay as $\Delta \tau_{AMI}= 25\Delta t=0.25$ s. 
To initialize the delay vector for the MAPSR method, we use a common difference of $\Delta \tau = 0.20$ s$\sim O (\Delta \tau_{AMI})$. 
For example, for 3-dimensional phase space $\vec{\tau}_{init}=[0.00, 0.20, 0.40]$ s$= \Delta \tau [0,1,2]$. 
The delay vector for each of the six cases (with different noise levels) is initialized as a $d$ dimensional vector $\vec{\tau}=\Delta \tau [0,1,\ldots,d-1]$ here, due to the time series being univariate, the dimension of the delay vector $D$ is the same as that of $d$; i.e., $D=d$. 
Each case is tested by varying $d$ from 1 to 5 in steps of 1. 
Linear interpolation is used to obtain the time series of the reconstructed state vector from the input time series based on the delay vector $\vec{\tau}$.

Training of the neural ODE is performed on 100 trajectories (batch size = 100) randomly chosen from the reconstructed attractor. 
Here, each trajectory is of 0.5 s duration; i.e., batch time is $0.5$ s or $t_{batch}=0.5$ s. 
A new batch is chosen for each training iteration. 
The loss is computed using Eq. (\ref{eq:mapsr_loss}). 
The learning rate for $W$ is chosen as $\alpha_{W}=10^{-3}$ and for $\vec{\tau}$ as $\alpha_{\vec{\tau}}=10^{-5}$. 
The training of the neural ODE is performed for fixed 20,000 iterations (it is observed but not shown here that within 20,000 iterations the loss function converges to a steady value and the components of the delay vector reach a steady value refer Fig. \ref{fig:harmonic_1}(b)). 
The average loss for the last 100 iterations of training is shown in Fig. \ref{fig:harmonic_1}(a). 
The figure shows that the training loss is minimum for the time series with $0\%$ noise and increases with the noise level. 
The minimum of the loss occurs for the phase space of dimension two, for all the time series with different noise levels.  
This might seem contradictory to the expected behavior for the loss function; that is, on increasing the dimension, the loss should decrease until trajectories get well resolved and should stay low afterwards.
The reasons for this observed behavior are discussed in \ref{subsec:analysis}.

The evolution of the delay values for the phase space with optimal dimension (with minimum loss) is shown in Fig. \ref{fig:harmonic_1}(b). 
The optimal dimensions predicted by the MAPSR method is 2 for all the time series with different noise levels. 
As discussed before, the figure shows the evolution of the delay vector for different time series with the delay vector initialized to $\Delta \tau[0, 1]$. For all the time series with different noise levels, the first delay value is zero and always maintained at zero throughout the training in accordance with Eq. \ref{eq:mapsr_tau_update}. 
Hence, the lines showing the evolution of the first delay value, which is zero, overlap for all the cases. 
For the clean time series without noise, the second component of the delay vector approaches $0.25$ s, which is similar to the delay predicted by the AMI-FNN method. 
As the noise level increases, the predicted value of the second component of the delay vector is lower than $0.25$ s, i.e., the delay value stays closer to its initial value. 
Thus, we can see that the linear independence of the coordinates is not strictly enforced, but the linear model for the clean time series naturally leads to linearly independent coordinates.

The results from the MAPSR method are compared with the AMI-FNN method and PECUZAL method in Table \ref{tab:harmonic}. 
For a clean signal with the noise level of 0 $\%$ AMI-FNN predicts the dimension as 7 and delay vector with $\Delta \tau = 0.25$ s. 
Knowing the dynamics of the harmonic oscillator, the dimensions predicted are indeed quite high. 
The MAPSR and PECUZAL methods predict the dimension as 2 and estimate the same delay vector. 
The PECUZAL method is based on the concept of noise amplification and fails for the 0 $\%$ noise level; however, the addition of a minute level of noise solves the problem. 
On addition of the noise, the AMI-FNN method estimates the same $\Delta \tau = 0.25$ s and the dimension of 4 for all the cases. 
The MAPSR method estimates a dimension of 2 for all the noise levels, but the second component of the delay vector starts deviating from 0.25 s on increasing the noise. 
The PECUZAL method estimates the dimension of 2 for 1 $\%$ and 4 $\%$ noise level, 3 for 2 $\%$ noise level, and 4 for 3 $\%$ noise level. 
Thus, the MAPSR method estimates the same dimensions for all noise levels, whereas the AMI-FNN predicts the highest dimensions for all the situations; on the other hand, the PECUZAL method estimates a higher number of dimensions with the increase in the noise level. 
Here we observe that the MAPSR has estimated the least dimensions compared to the AMI-FNN and the PECUZAL methods. 
Also, the estimated delay value of $0.25$ s for clean data agrees with both of the methods. 
The delay value gets less modified as the noise level increases.

The comparison of the phase portrait reconstructed using the true input time series and the trajectories predicted using a model trained with the MAPSR method are shown in Fig. \ref{fig:harmonic_phase_portrait}. 
For 0\% noise level, Fig. \ref{fig:harmonic_phase_portrait}(a), the true and predicted trajectories overlap exactly with each other, and the attractor shape is nearly circular. 
For other noise levels, Fig. \ref{fig:harmonic_phase_portrait}(b)-(f), the model is able to capture the attractor of an elliptical shape.

\subsection{Application of the MAPSR method to univariate time series from the Lorenz system}

The MAPSR method is further tested with the time series obtained from the Lorenz system \cite{paper:lorenz_1963}. 
The $x$ time series of the Lorenz system
\begin{equation}
    \begin{split}
    \dot{x} &=\sigma(y-x)\\
    \dot{y} &=x(\rho-z)-y\\
    \dot{z} &=xy-\beta z
    \end{split}
    \label{eq:Lorenz}
\end{equation}
is selected for the analysis, with parameters $(\rho, \sigma, \beta)= (28, 10, 8/3)$, integrated with a DOPRI5 \cite{book:DOPRI5_1993} solver at a step size $\Delta t=0.01$ s. Similar to the harmonic oscillator, noise is added to the $x(t)$ time series of the Lorenz system to test the method for six noise levels, from $0\%$ to $5\%$, with a step size of $1\%$.

For the MAPSR method, a neural ODE is used with 3 hidden layers and 4 weight matrices. 
Each hidden layer has 50 nodes. 
This configuration of neural ODE is arbitrarily chosen (other modeling techniques can be used in place of neural ODE).
For nonlinearity, we used the $tanh$ activation function.
The state of the last hidden layer is linearly mapped to the output vector $\dot{\vec{x}}$ of dimension $d$ without any nonlinearity.

We randomly choose 300 different initial conditions (i.e., batch size = 300), and the model is used for predicting the next 25 states with a time step of $\Delta t=0.01$ s or for the time of 0.25 s (i.e., batch time = 0.25 s). 
The weight matrix and delay vector are trained with a learning rate of $\alpha_{W}=10^{-3}$ and $\alpha_{\vec{\tau}}=10^{-5}$ respectively. 
Training is performed with a fixed 20,000 iterations.

\begin{figure*}
    \centering\linespread{2}
    \includegraphics[width=\textwidth]{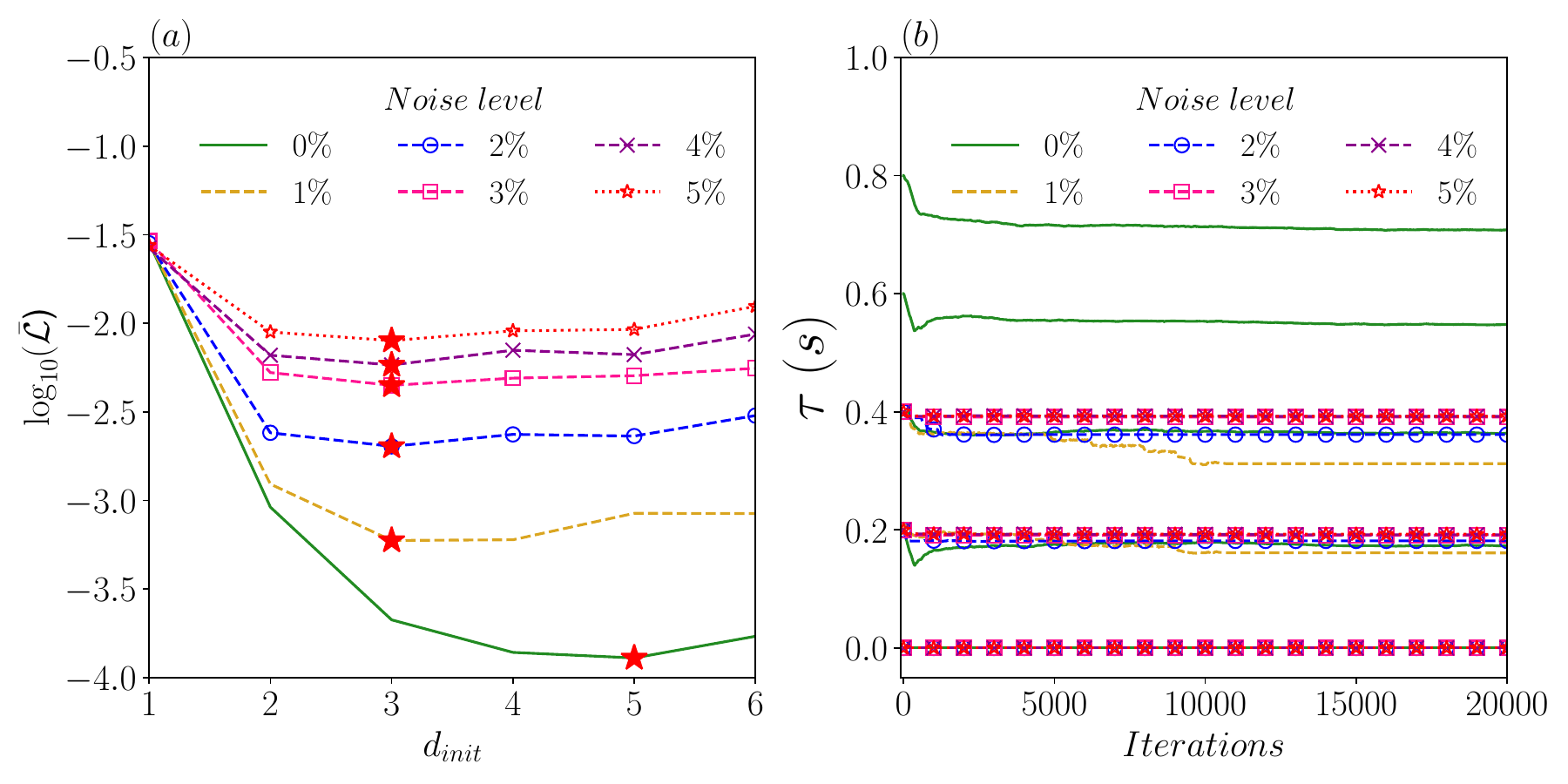}
    \caption{Parameters of phase space reconstruction for the time series data from Lorenz system with different noise levels using MAPSR method. 
    (a) The plot of the average loss ($\bar{\mathcal{L}}$) in $\log$ scale for different initial dimensions $d_{init}$. 
    The point with minimum loss is shown with ({\color{red}$\star$}) for each noise level. 
    For 0\% noise, the loss is minimum for a dimension of 5. 
    For the remaining noise levels, the method estimates the dimension as 3. 
    (b) Shows the evolution of the delay vector for optimal dimension with training iterations for time series with different noise levels. 
    The lines showing the evolution of the first three delay values (approximately) overlap, for the 0\% and 2\%  and also for 3\% to 5\% noise levels.}
    \label{fig:Lorenz_cmp}
\end{figure*}

\begin{table*}[ht]
    \centering\linespread{2}\renewcommand{\arraystretch}{1.5}
    \begin{tabular}{|p{3cm}||p{2cm}|c|p{7cm}|}
    \hline
    Case & Method & Dimension & Delay (s)\\
    \hline\hline
    \multirow[c]{3}{*}{Noise level: 0\%} & AMI-FNN & 3 & [0.0, 0.17, 0.34] \\
     & MAPSR & 5 & [0.0, 0.17, 0.36, 0.55, 0.71] \\
     & PECUZAL & 3 & [0.0, 0.17, 0.09] \\
     \hline
    \multirow[c]{3}{*}{Noise level: 1\%} & AMI-FNN & 5 & [0.0, 0.17, 0.34, 0.51, 0.68] \\
     & MAPSR & 3 & [0.0, 0.16, 0.31] \\
     & PECUZAL & 4 & [0.0, 0.17, 0.9, 0.74] \\
     \hline
    \multirow[c]{3}{*}{Noise level: 2\%} & AMI-FNN & 5 & [0.0, 0.17, 0.34, 0.51, 0.68] \\
     & MAPSR & 3 & [0.0, 0.18, 0.36] \\
     & PECUZAL & 4 & [0.0, 0.18, 0.88, 0.71] \\
     \hline
    \multirow[c]{3}{*}{Noise level: 3\%} & AMI-FNN & 6 & [0.0, 0.17, 0.34, 0.51, 0.68, 0.85] \\
     & MAPSR & 3 & [0.0, 0.19, 0.39] \\
     & PECUZAL & 4 & [0.0, 0.18, 0.87, 0.44] \\
     \hline
    \multirow[c]{3}{*}{Noise level: 4\%} & AMI-FNN & 7 & [0.0, 0.19, 0.38, 0.57, 0.76, 0.95 1.14] \\
     & MAPSR & 3 & [0.0, 0.19, 0.39] \\
     & PECUZAL & 4 & [0.0, 0.19, 0.85, 0.38] \\
     \hline
    \multirow[c]{3}{*}{Noise level: 5\%} & AMI-FNN & 7 & [0.0, 0.19, 0.38, 0.57, 0.76, 0.95 1.14] \\
     & MAPSR & 3 & [0.0, 0.19, 0.39] \\
     & PECUZAL & 4 & [0.0, 0.18, 0.8, 0.98] \\
     \hline
    \end{tabular}
    \caption{Embedding dimension and delay values estimated for the $x$ time series of Lorenz system with noise levels from 0 to 5\% in step of 1\%. 
    The estimates from the MAPSR method are compared with the AMI-FNN and PECUZAL methods.}
    \label{tab:lorenz}
\end{table*}

\begin{figure*}
    \centering\linespread{2}
    \includegraphics[width=\textwidth]{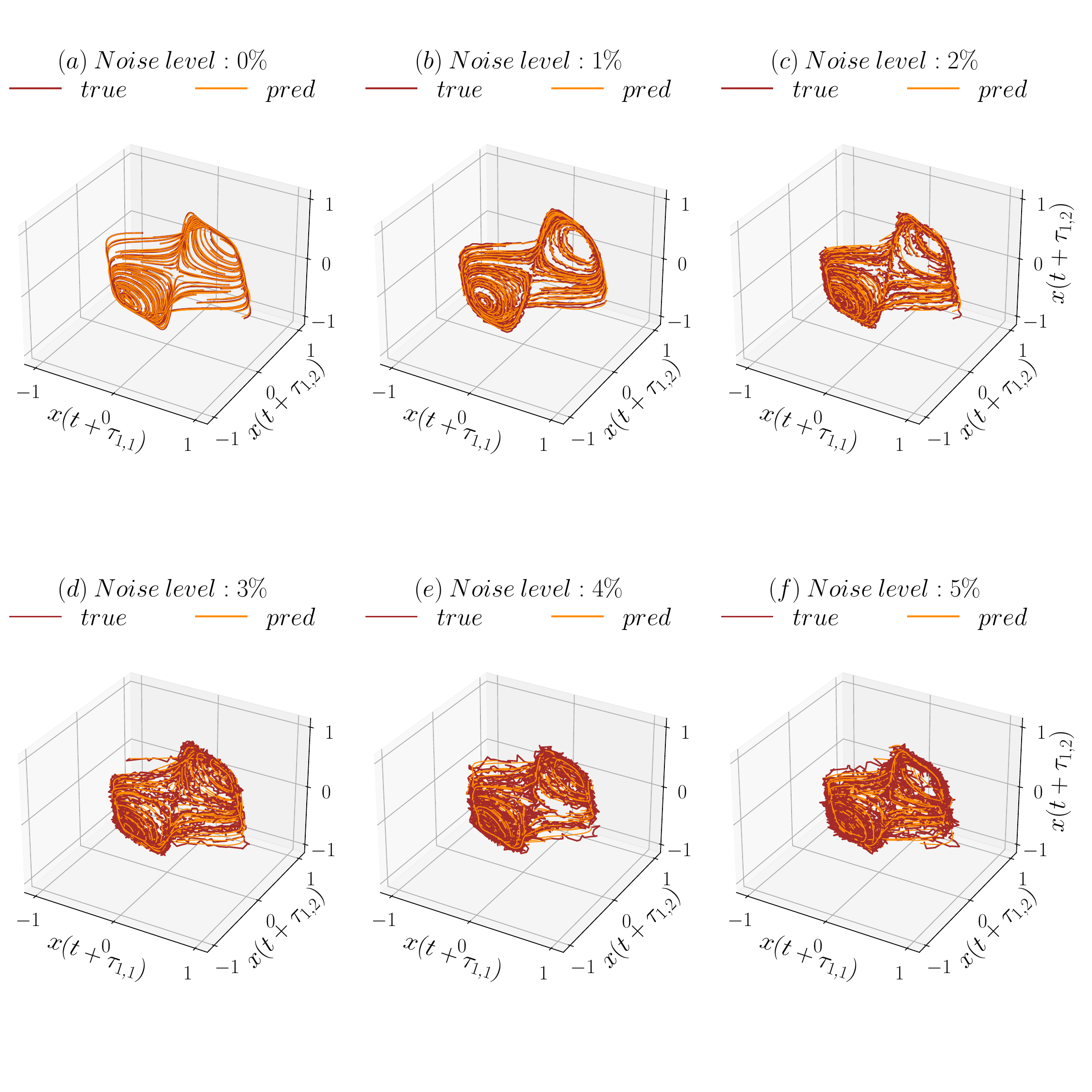}
    \caption{Comparison of the true trajectories reconstructed from the $x$ time series of the Lorenz system with the predicted trajectories using MAPSR method for six different noise levels 0\%-5\% with neural ODE as the model.}
    \label{fig:Lorenz_phase_diag}
\end{figure*}

 The AMI-FNN method estimates the $\Delta \tau_{AMI}$ as $0.17$ s and the dimension as 3 for the Lorenz system without noise. 
 Thus, for the MAPSR method, the delay vector is initialized with a common difference of $\Delta \tau = 0.2$ s$\sim O(\tau_{AMI})$. 
 Each time series is tested for the initial dimension from 1 to 6. 
 The average loss (in $\log$ scale) for the last 100 iterations for all the time series with different embedding dimensions is shown in Fig. \ref{fig:Lorenz_cmp}(a). 
 The configuration with minimum loss is marked with a star ({\color{red} $\star$}) and is chosen as optimal embedding. 
 For a clean time series with 0\% noise, MAPSR predicts the optimal embedding dimension as 5, whereas for all other noise levels, the embedding dimension of 3 leads to the minimum loss. 
 The evolution of the components of the delay vector during the training for optimal configuration is shown in Fig. \ref{fig:Lorenz_cmp}(b). 
 Similar to the harmonic oscillator, the delay values adjust for the first few iterations and then attain a steady value except for the time series with 1\% noise level for which delays update till (approx.) 10,000 iterations, after which they become steady. 
 The first three delay values (approx.) attain the same value for 0\% and 2\%  and also for 3\% to 5\% noise levels. 
 The exact values for the delay are given in Table \ref{tab:lorenz}.      

The dimensions and delays estimated for time series with different noise levels are given in Table \ref{tab:lorenz}. 
For clean time series, the AMI-FNN and PECUZAL predict the embedding dimension as 3, whereas the MAPSR method predicts the dimension as 5, which is indeed quite large. 
The first two delay values for all three methods are the same. 
For higher noise levels, the embedding dimension estimated with the AMI-FNN method increases, whereas PECUZAL and MAPSR method estimates the same embedding dimension as 4 and 3, respectively. 
The first two delay values estimated by all the methods are very close. 
For the third delay value, only the estimates from AMI-FNN are closer to MAPSR. 
For the fourth delay value, there is a significant difference in the values predicted by AMI-FNN and PECUZAL methods. 
The higher dimensions predicted by the AMI-FNN method can be attributed to increased false nearest neighbors (FNN) detected due to noise. 
The MAPSR predicts the embedding dimension of 3, which might be due to the estimation of trajectories using neural ODE, which is smooth and reduces the effect of noise on the embedding dimension estimation.
We can see here that increasing the noise level decreases the number of dimensions required for the modeling as compared to the clean data.
The same has been conjectured by \cite{paper:garland_2015}, which states that "a full formal embedding, although mandatory for detailed dynamical analysis, is not necessary for the purposes of prediction," this especially holds for noisy data and has been discussed in \cite{paper:bradly_kantz_2015}.
 
The three-dimensional phase portrait of the reconstructed attractor using $x$ time series from the Lorenz attractor ($true$) is compared to the trajectories predicted using the neural ODE model ($pred$) in Fig. \ref{fig:Lorenz_phase_diag} for different noise levels. 
For the clean time series Fig. \ref{fig:Lorenz_phase_diag}(a), the true and predicted time series closely overlap. 
For other noise levels Fig. \ref{fig:Lorenz_phase_diag}(b)-(f), the true trajectories are distorted due to noise, but the trajectories predicted using the MAPSR method are smooth and able to follow true noisy trajectories. 

\begin{figure*}
    \centering\linespread{2}
    \includegraphics[width=\textwidth]{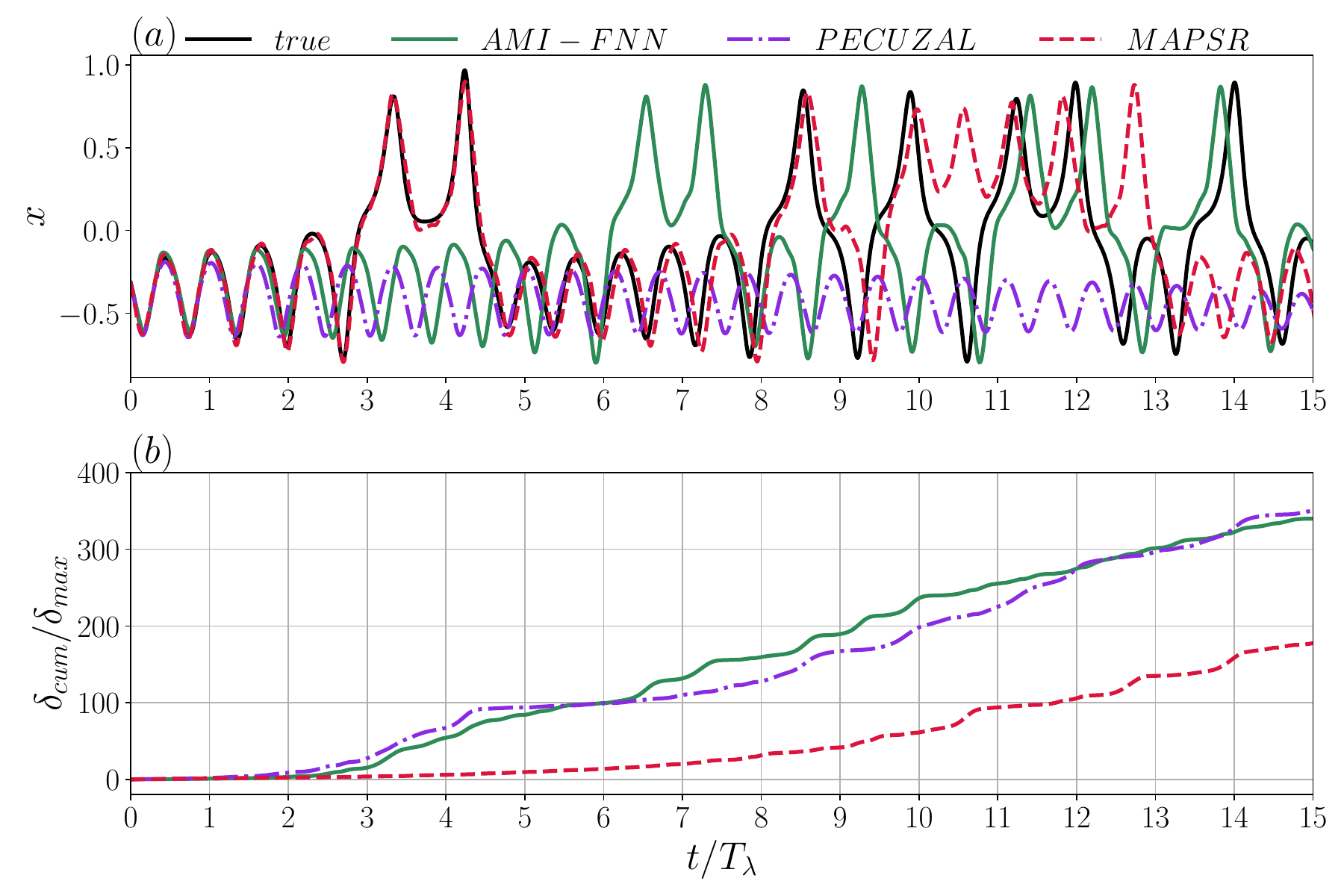}
    \caption{(a) Comparison of the true $x$ time series of the Lorenz system (Eq. \ref{eq:Lorenz}) with the time series predicted using the neural ODE trained on the trajectories reconstructed using AMI-FNN, PECUZAL and MAPSR methods. 
    The predicted time series follows the true time series close to 2-3$T_{\lambda}$ for AMI-FNN and PECUZAL method and for 8-9$T_{\lambda}$ for MAPSR method. (b) The evolution of the cumulative normalized deviation ($\delta_{cum}/\delta_{\max}$) with normalized time ($t/T_{\lambda}$).}
    \label{fig:Lorenz_loss_evol}
\end{figure*}

To compare the prediction horizon of the MAPSR method with the AMI-FNN and PECUZAL methods, we train the neural ODE on the trajectories reconstructed using AMI-FNN and PECUZAL methods independently.
Here, we used the same set of hyperparameters for neural ODE except for the number of input and output nodes, which are PSR method-specific. 
The neural ODE is used with 50 nodes in each of the 3-hidden layers. 
We used the clean time series of $x$ from the Lorenz system for PSR using AMI-FNN and PECUZAL method.
The model is trained to predict for the duration of $T_{R}\sim 1.4 T_{\lambda}$, where $T_{\lambda}=1/\lambda$ is the Lyapunov time scale and $\lambda$ is the Lyapunov exponent. 
The MAPSR estimates a five-dimensional phase space (this estimated dimension is high due to large $T_R$). 
The comparison of the true time series of $x$ obtained using Eq.~(\ref{eq:Lorenz}) with the time series of $x$ estimated using the AMI-FNN, PECUZAL and MAPSR method is shown in Fig. \ref{fig:Lorenz_loss_evol}(a). 
The predicted time series follows the true time series for the duration close to $2T_{\lambda}-3T_{\lambda}$ for AMI-FNN and PECUZAL method and $8T_{\lambda}-9T_{\lambda}$ for MAPSR method. 
The normalized cumulative deviation $(\delta_{cum}/\delta_{\max})$ of the predicted time series from the true time series is shown in Fig. \ref{fig:Lorenz_loss_evol}(b), here, $\delta_{cum}(t)=\sum_{0\leq t_i\leq t}|x_{pred}(t_i)-x_{true}(t_i)|$ and  $\delta_{max}=\max(x_{true})-\min(x_{true})$. 
The normalized cumulative deviation for AMI-FNN and PECUZAL method shows similar trend where for MAPSR method $\delta_{cum}$ grows gradually compared to the AMI-FNN and PECUZAL method. 

One can improve the predictions of the neural ODE trained using trajectories from AMI-FNN and PECUZAL methods by adjusting neural ODE configuration and hyperparameters. 
We find that MAPSR method which uses delay embedding optimized for considered neural ODE configuration and set of hyperparameters performs better than AMI-FNN and PECUZAL method.

\subsection{Application of the MAPSR method to univariate time series from a turbulent combustor}

After these two initial theoretical examples, we demonstrate MAPSR in the following on experimental data: time series data obtained from a turbulent combustor.
The experiments conducted on the turbulent combustor aim to study the transitions in thermoacoustic systems \cite{paper:unni_sujith_2015}. 
Thermoacoustic systems involve the interaction of the heat source and the acoustic field within the confining chamber\cite{book:sujith_2021}. 
The positive feedback between the acoustics and the heat source can lead to high amplitude limit cycle oscillations (LCO) and is well known as thermoacoustic instability in the field of gas turbines and rocket engines \cite{book:sujith_2021}. 
To study how the dynamical state of the thermoacoustic system changes with the airflow rate (control parameter), the experiments are conducted at a constant fuel (liquefied petroleum gas: butane $40\%$ and propane $60\%$) flow rate of 28 SLPM (standard liter per minute) and by quasi-steadily varying the air flow rate from 448 SLPM to 878 SLPM in steps of 28 SLPM. 
For each mass flow rate of air, $3$ s long time series of pressure fluctuations $p'$ and heat release rate $\dot{q}$ are recorded at a sampling rate of 10 kHz. 
For an air flow rate near 448 SLPM, the time series of $p'$ shows chaotic behavior \cite{paper:nair_chaos_2013, paper:tony_2015}, and near 878 SLPM shows the dynamical regime of limit cycle oscillations (LCO). 
The detailed description of the experiment is reported by \citet{paper:unni_sujith_2015}.
The transition from chaos to LCO occurs via intermittency \cite{paper:nair_thampi_sujith_2014}. 
Here we test the proposed methodology to obtain delay embedding and model the different dynamical regimes of the turbulent thermoacoustic system.  

\begin{figure*}
    \centering\linespread{2}
    \includegraphics[width=\textwidth]{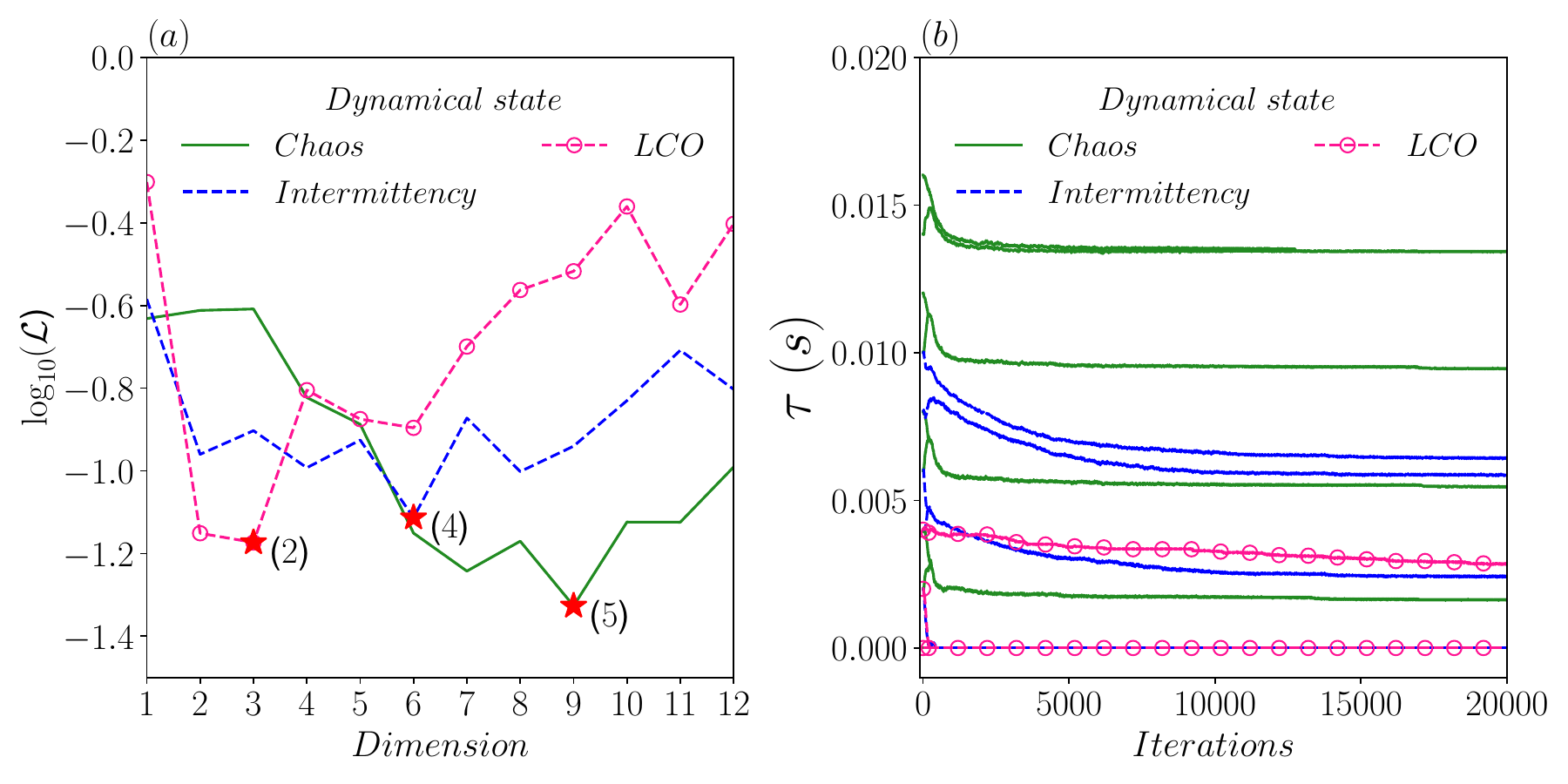}
    \caption{Application of the MAPSR method to the time series data from the turbulent combustor. 
    (a) The variation of the average loss ($\mathcal{\bar{L}}$) for different initial dimensions $d_{init}$ in $log$ scale. 
    The point with minimum loss is shown with ({\color{red}$\star$}) for each dynamical regime. 
    The bracket next to ({\color{red}$\star$}) shows the dimensions ($d_{final}$) estimated using MAPSR method. 
    For the time series of chaos, the training starts with three-dimensional phase space, which shrinks to two-dimensional space during training. 
    For the time series data with intermittency and LCO, the initial dimension of  6 and 9 shrinks to 4 and 5, respectively. 
    (b) Shows the evolution of delay vector with training iterations. 
    The delays that have moved closer get merged and can be observed for the first few iterations (The enlarged view is shown in Fig. \ref{fig:exp_zoom}). 
    Further, the delay values converge with iterations and stay nearly constant.}
    \label{fig:exp_both}
\end{figure*}

\begin{figure*}
    \centering\linespread{2}
    \includegraphics[width=0.6\textwidth]{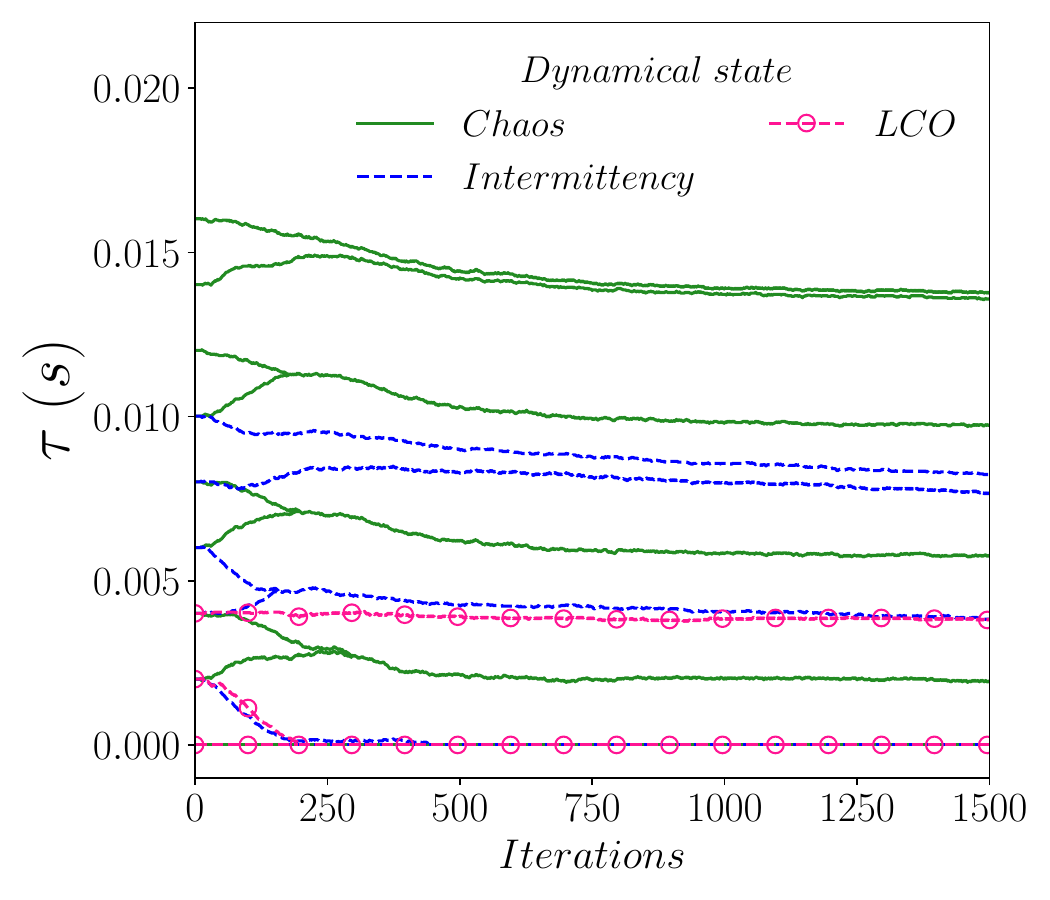}
    \caption{Variation of delay vector during the training with experimental data from different dynamical regimes. 
    The plot shows an enlarged view of Fig. \ref{fig:exp_both}(b) for the first few iterations. 
    The delay values merge if they are closer than a threshold $\tau_{th}=\Delta t/2$. 
    For the chaos ({\color{ForestGreen}---}), the delay vector of initial length 9 shrinks to 5 during the training. 
    The delay vector with an initial length of 6 and 3 shrinks to 4 and 2  for the regime of intermittency ({\color{blue} - - -}) and LCO  ({\color{Magenta} $- \ominus -$}) respectively.}
    \label{fig:exp_zoom}
\end{figure*}

The time series data from turbulent combustor in three different regimes, a) chaos, b) intermittency, and c) LCO, is used to assess the behavior of MAPSR on real-world time series data. 
The AMI-FNN method predicts the dimension of the phase space as 5 and $\Delta \tau_{AMI} = 0.26\times10^{-2}$ s for the regime of chaos. 
Thus, the delay vector for the MAPSR method is initialized with a common difference of $\Delta \tau = 0.2\times10^{-2}$ s$\sim O(\Delta \tau_{AMI})$ with the first delay value equal to zero and $d_{init}$ delay components. 
The MAPSR method compares the training loss for $d_{init}$ = 1 to 12. 
The neural ODE used for modeling the dynamics has 3 hidden layers and 4 weight matrices. 
The learning rate for the weight matrices is initially maintained at $10^{-3}$, which is smoothly changed to $10^{-4}$ while training. 
The delay vector is trained with a fixed learning rate of $10^{-6}$. 
Using the delay vector and time series data from experiments, the attractor is reconstructed using linear interpolation. 
Randomly 60 points are chosen on the attractor as the initial conditions, i.e., batch size = 60, and neural ODE is integrated for $0.72T$ s; i.e., batch time = $0.72 T$ s, with each of these as initial conditions. Here, $T$ s is the period of the oscillations during the regime of LCO. 
The loss is computed using Eq.~(\ref{eq:mapsr_loss}) with $L_1$-norm. 

\begin{table*}
    \centering\linespread{2}\renewcommand{\arraystretch}{1.5}
    \begin{tabular}{|p{3cm}||p{2cm}|c|p{7cm}|}
    \hline
    Case & Method & Dimension & Delay $(\times 10^{-2})$ s\\
    \hline\hline
    \multirow[c]{3}{*}{Chaos} & AMI-FNN & 5 & [0.0  , 0.26, 0.52, 0.78, 1.04] \\
     & MAPSR & 5 & [0.0    , 0.1623, 0.5448, 0.9474, 1.3439] \\
     & PECUZAL & 5 & [0.0  , 0.29, 0.14, 0.22, 0.07] \\
     \hline
    \multirow[c]{3}{*}{Intermittency} & AMI-FNN & 5 & [0.0  , 0.24, 0.48, 0.72, 0.96] \\
     & MAPSR & 4 & [0.0    , 0.2425, 0.5831, 0.6434] \\
     & PECUZAL & 6 & [0.0  , 0.24, 0.12, 0.18, 0.48, 0.35] \\
     \hline
    \multirow[c]{3}{*}{LCO} & AMI-FNN & 5 & [0.0 , 0.2, 0.4, 0.6, 0.8] \\
     & MAPSR & 2 & [0.0    , 0.2845] \\
     & PECUZAL & 5 & [0.0  , 0.21, 0.11, 0.33, 0.27] \\
     \hline
    \end{tabular}
        \caption{Embedding dimension and delay values for three dynamical regimes of the turbulent combustor a) chaos, b) intermittency, c) LCO, estimated with different phase space reconstruction methods, i.e., AMI-FNN, MAPSR, and PECUZAL.}
        \label{tab:exp}
    \end{table*}

The loss incurred after training for dynamical regimes of chaos, intermittency, and LCO is shown in Fig. \ref{fig:exp_both}(a). 
The horizontal axis shows the initial dimensions ($d_{init}$), i.e., at the start of training. 
The dimension at which loss is minimum is shown with a star mark accompanying the value within bracket $(d_{final})$, where $d_{final}$ is the optimal embedding dimensions after the training. 
For all three dynamical regimes, $d_{init}$ and $d_{final}$ are different due to the merging of the delays as discussed in Sec. \ref{sec:MAPSR_description}. 
For the regime of chaos, the MAPSR method estimates the dimension of the delay vector as $d_{final}=5$, which was initialized to $d_{init}=9$ before training. 
Similarly, for the dynamical regimes of intermittency and LCO, the MAPSR method estimates the dimension as 4 and 2, which were initialized to 6 and 3, respectively. 
The development of the delay vector for the optimal cases marked by the star in Fig. \ref{fig:exp_both}(a) is shown in Fig. \ref{fig:exp_both}(b). 
The first delay value for all the cases is zero. 
The merging of the delay occurs during the first few iterations, and a zoomed view is shown in Fig. \ref{fig:exp_zoom}. 
For the regime of chaos the initial delay vector is $\Delta\tau[0,1,2,3,4,5,6,7,8]$. 
During training, the first delay component stays zero. 
The second-third, fourth-fifth, sixth-seventh, and eighth-ninth components merge to give a five-dimensional delay vector. 
For the regime of intermittency, the initial delay vector is $\Delta\tau[0,1,2,3,4,5]$ whose first-second and third-fourth components merge and give four-dimensional phase space. 
Similarly, for the regime of LCO, the initial delay vector is $\Delta\tau[0,1,2]$, whose first and second component merge and give a two-dimensional phase space.

\begin{figure*}
    \centering\linespread{2}
    \includegraphics[width=1.0\textwidth]{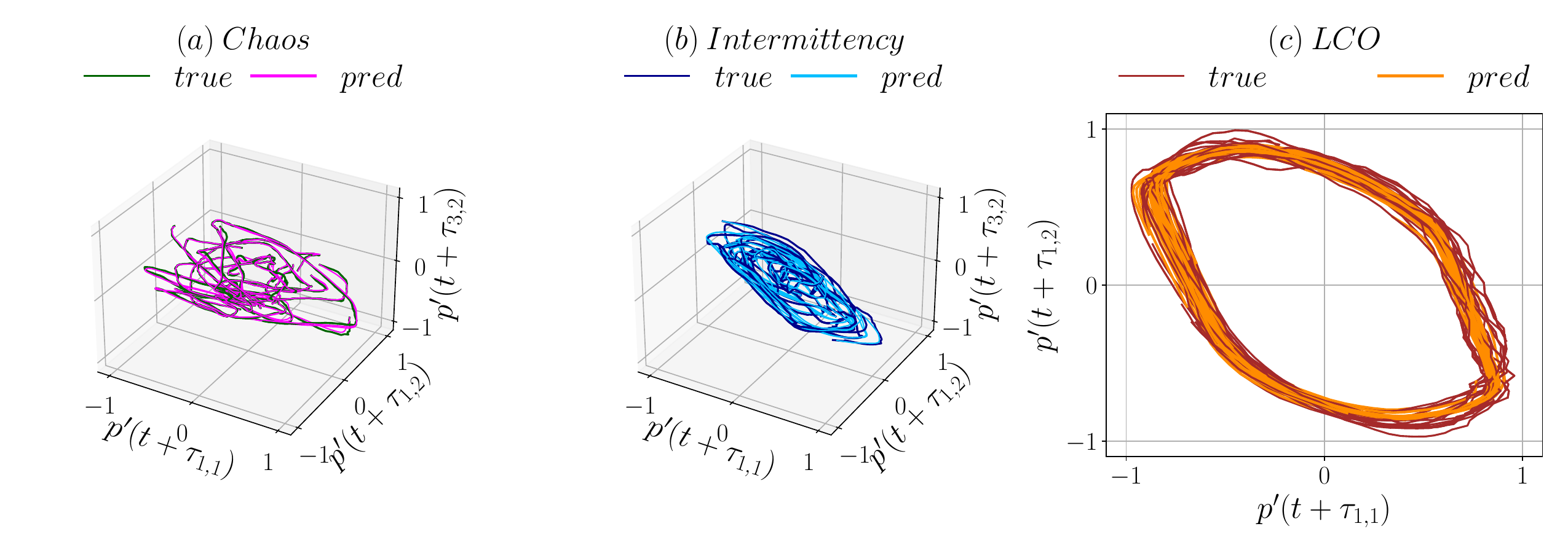}
    \caption{Comparison of the true trajectories reconstructed from the univariate time series of $p'$ with the trajectories predicted using MAPSR method using neural ODE as a model.}
    \label{fig:exp_phase_uni}
\end{figure*}

\begin{figure*}
    \centering\linespread{2}
    \includegraphics[width=1.0\textwidth]{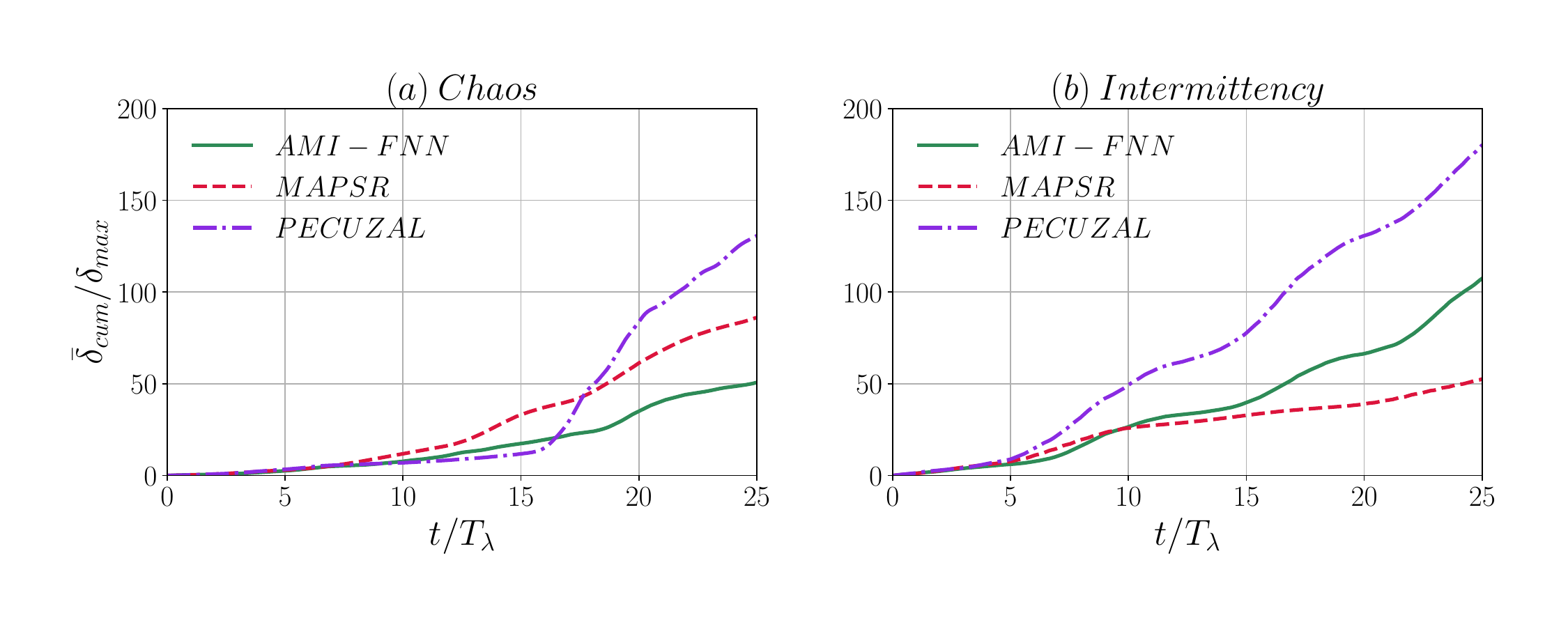}
    \caption{Comparison of the evolution of the normalized average cumulative deviation ($\bar{\delta}_{cum}/\delta_{\max}$) of the predicted time series of $p'$ for AMI-FNN, MAPSR and PECUZAL methods with normalized time ($t/T_{\lambda}$) for the dynamical regime of (a) chaos, (b) intermittency.}
    \label{fig:exp_loss_uni}
\end{figure*}

The trajectories reconstructed using the true time series of $p'$ from the turbulent combustor are compared with the predicted time series using the MAPSR method with the neural ODE model in Fig.\ref{fig:exp_phase_uni}. 
For the dynamical regimes of chaos, intermittency, and LCO, the predicted trajectories closely follow the true trajectories reconstructed using true time series of $p'$. 
For the dynamical regimes of chaos and intermittency, Fig. \ref{fig:exp_phase_uni}(a) and Fig.\ref{fig:exp_phase_uni}(b), only the first three delay coordinates are used for visualization of the attractor in three-dimensional space. 
The delay embedding for the dynamical regime of LCO is estimated to be two-dimensional using the MAPSR method (Fig.~\ref{fig:exp_phase_uni}(c)).

The delay vector estimated with AMI-FNN, MAPSR, and PECUZAL methods for the data obtained from turbulent combustor are given in Table \ref{tab:exp}. 
For the dynamical regime of chaos, all the methods estimate the embedding dimension as 5, but the estimated delay values are different. 
For the dynamical regime of intermittency, the MAPSR method estimates the least dimension as 4, whereas AMI-FNN and PECUZAL estimate the dimension as 5 and 6, respectively. 
Also, for the regime of intermittency, the estimates of the first two delay values are approximately the same, i.e., $[0.0,\:0.24]\times10^{-2}$ s for all the methods. 
For the dynamical regime of LCO, AMI-FNN, and PECUZAL methods estimate the same embedding dimension of 5, whereas the MAPSR method estimates $2$ dimensional delay embedding, which is expected from LCO.
Here, we can see that for all dynamical regimes, the dimension estimated by the MAPSR is less than or equal to that predicted using other methods. 

Figure \ref{fig:exp_loss_uni} compares the prediction results for the AMI-FNN, MAPSR, and PECUZAL methods.
The evolution of the normalized average cumulative deviation ($\bar{\delta}_{cum}/\delta_{\max}$) with normalized time $t/T_\lambda$ for the dynamical regimes of chaos and intermittency are respectively shown in Fig. \ref{fig:exp_loss_uni}(a) and Fig. \ref{fig:exp_loss_uni}(b). 
Here, $\bar{\delta}_{cum}(t)$ is the cumulative deviation upto to time $t$ averaged over the batch, and $T_\lambda$ is the Lyapunov time scale for the time series of $p'$. 
For the dynamical regime of chaos, $(\bar{\delta}_{cum}/\delta_{\max})$ grows similarly for all the methods for nearly $7T_{\lambda}$.
After which $(\bar{\delta}_{cum}/\delta_{\max})$ grows faster for the MAPSR method and slowest for the PECUZAL method.
After $16T_{\lambda}$, the deviation for the PECUZAL method grows rapidly beyond other methods.
For the dynamical regime of intermittency, $(\bar{\delta}_{cum}/\delta_{\max})$ is same for all the methods upto $5T_{\lambda}$.
Further, the deviation grows fastest for the PECUZAL and slowest for AMI-FNN.
The deviation for MAPSR stays closer to AMI-FNN and drops below AMI-FNN after $10T_{\lambda}$. 

\begin{figure*}
    \centering\linespread{2}
    \includegraphics[width=\textwidth]{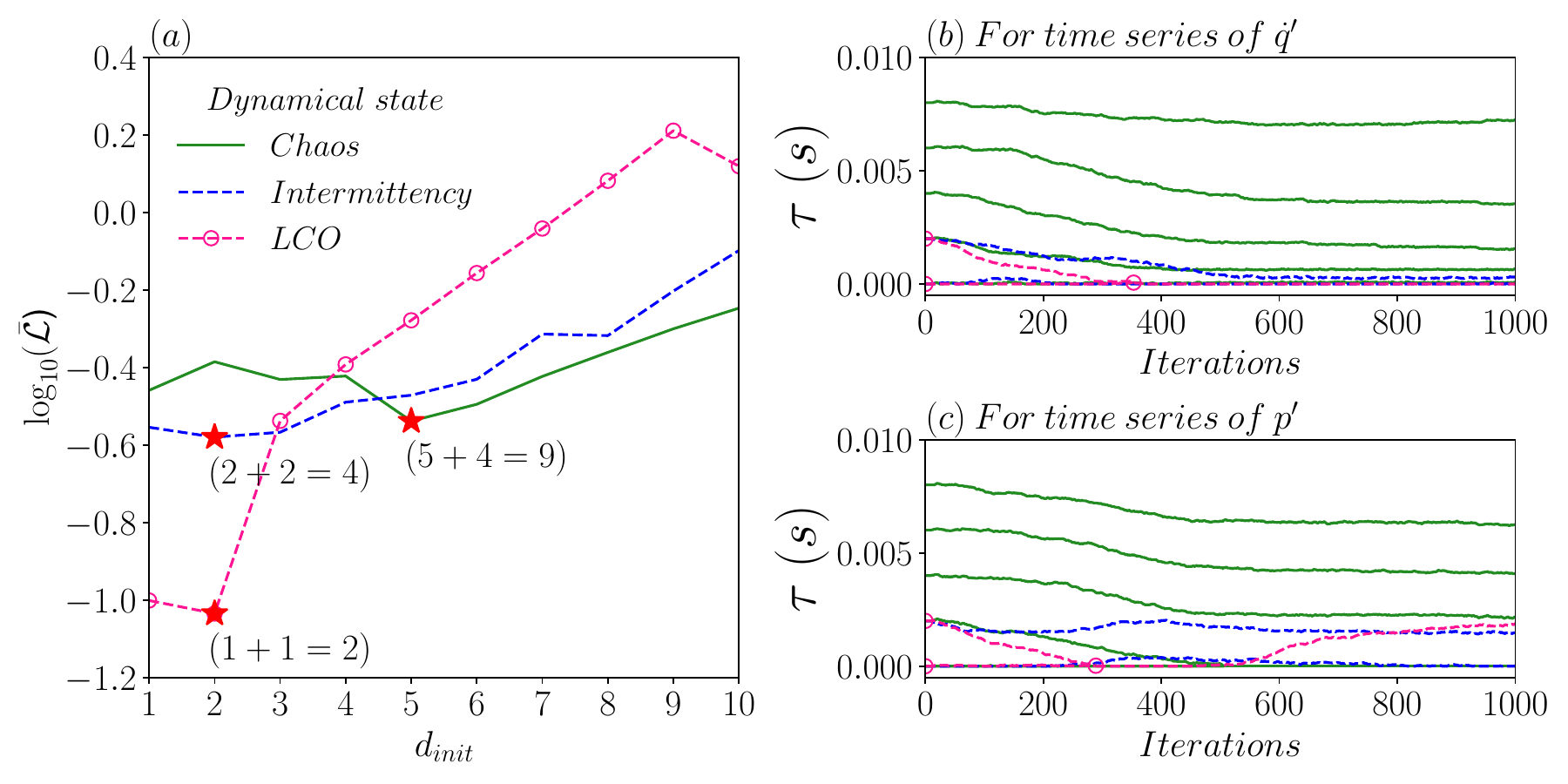}
    \caption{Variation of loss and delay vector for multivariate time series $(\dot{q}',\; p')$ from turbulent combustor. 
    $(a)$ The horizontal axis shows the initial dimension for each time series. Here there are two time series hence dimension of 4 on the horizontal axis means the initial embedding dimension is 4 for each time series hence net embedding dimension is, 4 + 4 = 8. 
    The points with minimal loss are marked by (${\color{red}\star}$). 
    For the chaotic time series ({\color{ForestGreen}---}), the delay vector of initial length 5 + 5 = 10 shrinks to 5 + 4 = 9 during the training. 
    The delay embedding with initial dimension of 2 + 2 = 4 stay as it is for intermittency ({\color{blue} - - -}) where as initial dimension of 2 + 2 = 4 shrinks to 1 + 1 = 2 for the regime of LCO  ({\color{Magenta} $- \ominus -$}). 
    $(b)-(c)$ Shows the evolution and merging of the delay values for the time series of $\dot{q}'$ and $p'$. 
    The same color and markers are used as Fig. $(a)$.}
    \label{fig:exp_multivariate}
\end{figure*}

\begin{table*}
    \centering\linespread{2}\renewcommand{\arraystretch}{2}
    \begin{tabular}{|p{3cm}||p{2cm}|c|p{8cm}|}
    \hline
Case & Method & Dimension & Delay $(\times 10^{-2})$ s\\
\hline\hline
\multirow[c]{2}{*}{Chaos} & MAPSR & 9 & \makecell[l]{\renewcommand{\arraystretch}{2}
    \begin{tabular}{lc}
    $\dot{q}'$ : [0.006 , 0.0643, 0.1445, 0.3748, 0.7351]\\
    $p'$ : [0.0    , 0.1835, 0.4128, 0.6031]\\
    \end{tabular}}
 \\\cline{2-4}
 & PECUZAL & 5 & \makecell[l]{\renewcommand{\arraystretch}{2}
                                \begin{tabular}{lc}
                                    $\dot{q}'$ : [0.0]\\
                                    $p'$ : [0.0  , 0.28, 0.14, 0.21]\\
                                \end{tabular}}
 \\\hline
\multirow[c]{2}{*}{Intermittency} & MAPSR & 4 & \makecell[l]{                                                                    \renewcommand{\arraystretch}{2}
                                        \begin{tabular}{lc}
                                            $\dot{q}'$ : [0.0    , 0.0163]\\
                                            $p'$ : [0.006 , 0.2363]\\
                                        \end{tabular}}
 \\\cline{2-4}
 & PECUZAL & 6 & \makecell[l]{
                            \renewcommand{\arraystretch}{2}
                                        \begin{tabular}{lc}
                                $\dot{q}'$ : []\\
                                $p'$ : [0.0  , 0.24, 0.12, 0.18, 0.48, 0.35]
                                \end{tabular}}
 \\\hline
\multirow[c]{2}{*}{LCO} & MAPSR & 2 & \makecell[l]{
                                        \renewcommand{\arraystretch}{2}
                                        \begin{tabular}{lc}
                                            $\dot{q}'$ : [0.0]\\
                                            $p'$ : [0.188]\\
                                        \end{tabular}}
 \\\cline{2-4}
 & PECUZAL & 5 & \makecell[l]{
                                        \renewcommand{\arraystretch}{2}
                                        \begin{tabular}{lc}
                                                $\dot{q}'$ : []\\
                                                $p'$ : [0.0  , 0.21, 0.11, 0.33, 0.27]\\
                                        \end{tabular}}
 \\\hline
\end{tabular}
        \caption{Embedding dimension and delay values for three dynamical regimes of the turbulent combustor a) chaos, b) intermittency, c) LCO, using multivariate time series data $[\dot{q}',p']$, estimated with different phase space reconstruction methods; i.e., MAPSR, and PECUZAL.}
        \label{tab:exp_multi}
\end{table*}

\subsection{Application of the MAPSR method to multivariate time series from the turbulent combustor}

The MAPSR is also tested with multivariate time series data $\vec{s} = [s_1,s_2]= [\dot{q}',\:p']$ from the turbulent combustor ($\dot{q}'$ is the heat release rate fluctuation which is mean subtracted $\dot{q}$). 
The delay vector is initialized with $2d_{init}$ components with $d_{init}$ components for each time series. 
Similar to the univariate case, for both the time series, delay values are initialized with a common difference of $\Delta \tau_{AMI}=0.2\times10^{-2}$ s with the first delay value as zero; e.g., for $d_{init}=2$, $\vec{\tau}=\Delta \tau[0,1,0,1]=[\tau_{1,1},\tau_{1,2},\tau_{2,1},\tau_{2,2}]$. 
For the application of the MAPSR method to multivariate time series, the same configuration of neural ODE and the learning rates used are the same as that of the univariate case. 

The variation of the loss with initial dimension $d_{init}$ is shown in Fig. \ref{fig:exp_multivariate}(a). 
For the three dynamical regimes of chaos, intermittency, and LCO, the loss is minimum for initial dimension $d_{init}$ of 2, 2, and 5 for each time series $\dot{q}'$ and $p'$, i.e., the net initial dimension $D_{init}$ of the delay vector is 4, 4, 10 for these three dynamical regimes, respectively. 
In Fig. \ref{fig:exp_multivariate}(a), next to this minima ({\color{red}$\star$}), the bracket shows the final dimension estimated using the MAPSR method; i.e., $(d_{1, final}+ d_{2, final}=D_{final})$. Thus, the MAPSR method respectively estimates $D_{final}$ of 9, 4, and 2 for the dynamical regime of chaos, intermittency, and LCO, respectively.

\begin{figure*}
    \centering\linespread{2}
    \includegraphics[width=\textwidth]{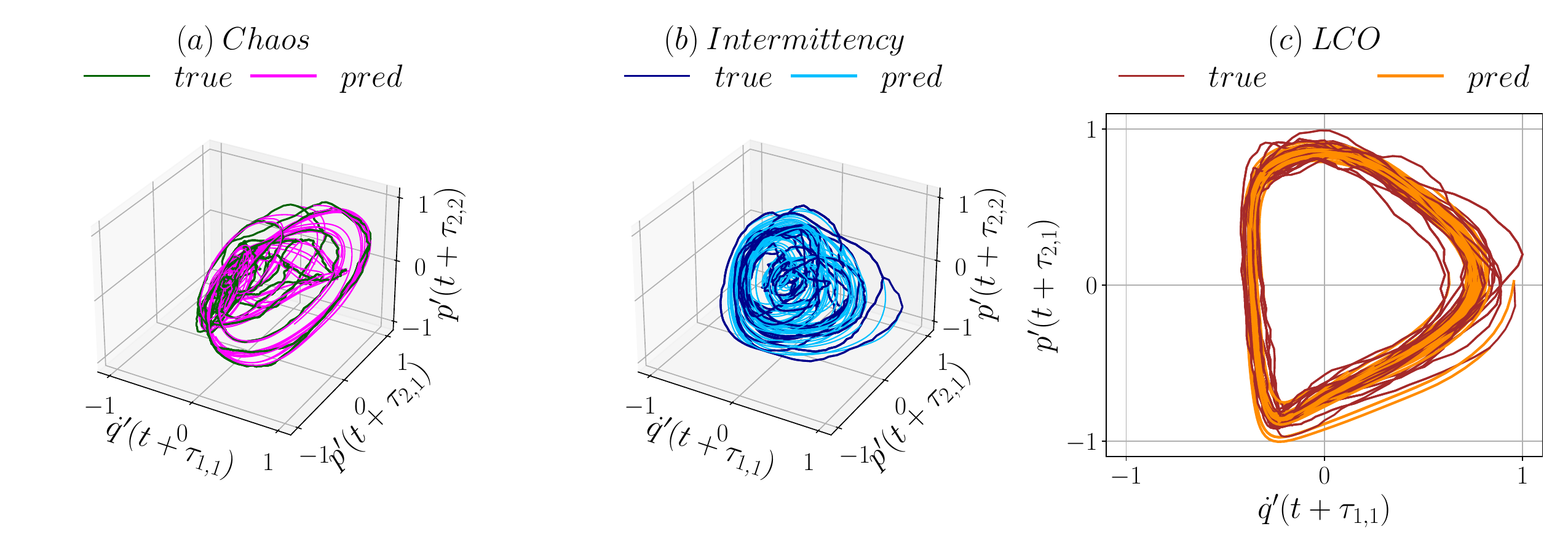}
    \caption{Comparison of the true trajectories reconstructed from the multivariate time series of ($\dot{q}'$, $p'$), with the trajectories predicted using MAPSR method using neural ODE as a model.}
    \label{fig:exp_phase_multi}
\end{figure*}

\begin{figure*}
    \centering\linespread{2}
    \includegraphics[width=1.0\textwidth]{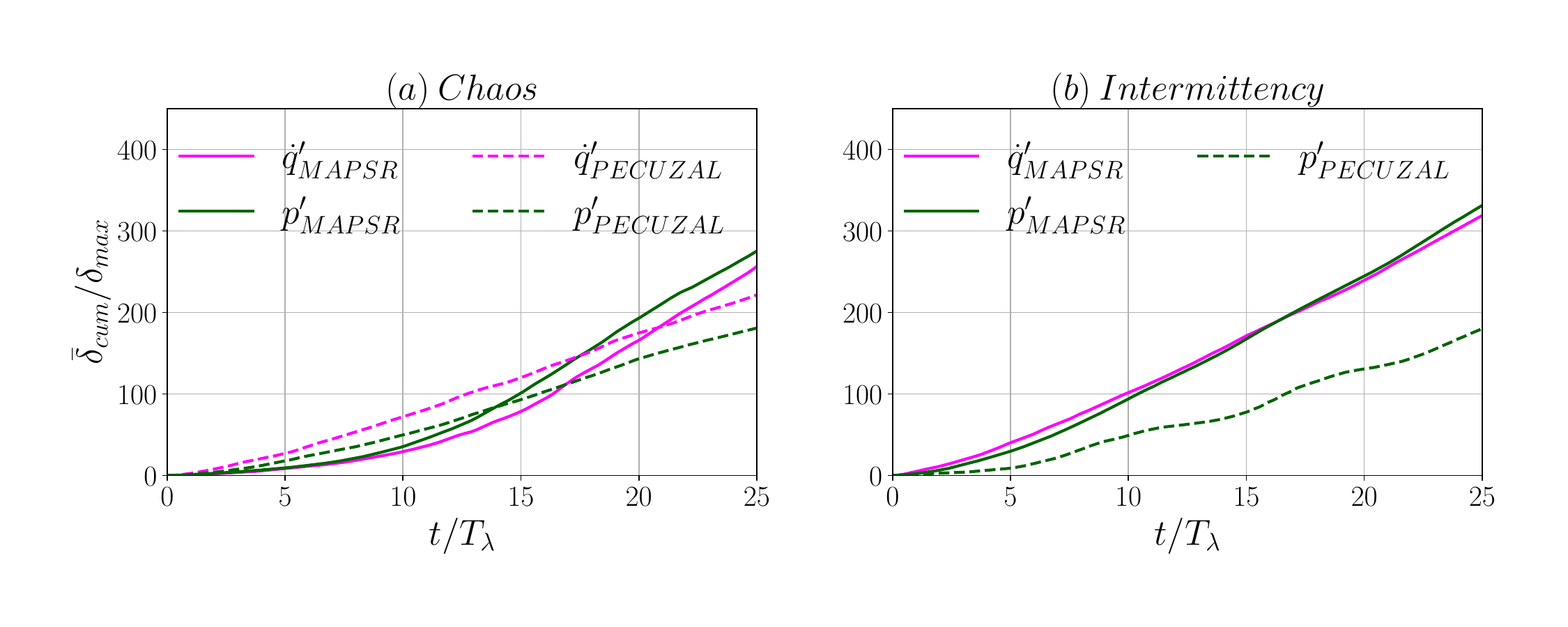}
    \caption{Evolution of the normalized average cumulative deviation ($\bar{\delta}_{cum}/\delta_{\max}$) of the predicted time series of $\dot{q}'$ and $p'$ for MAPSR and PECUZAL methods with normalized time ($t/T_{\lambda}$) for the dynamical regimes of (a) chaos, (b) intermittency. 
    Here, $T_{\lambda}$ is obtained using $p'$ time series, and $\delta_{\max}$ is obtained separately for individual time series. The PECUZAL method rejects $\dot{q}'$ time series, hence not shown for the case of intermittency.}
    \label{fig:exp_loss_multi}
\end{figure*}

The evolution of delay components for the time series of $\dot{q}'$ and $p'$ is shown in Figs. \ref{fig:exp_multivariate}(b) and \ref{fig:exp_multivariate}(c), respectively. 
For the dynamical regime of chaos, Fig. \ref{fig:exp_multivariate}(b) shows that initially, there are 5 components, and there is no merging of delay values for $s_1=\dot{q}'$ time series, whereas first-second delay values merge in case of $p'$ time series as shown in Fig. \ref{fig:exp_multivariate}(c). 
Hence, the final dimension of delay embedding for the regime of chaos is $D_{final}=5+4=9$. 
For the dynamical regime of intermittency, Fig. \ref{fig:exp_multivariate}(b) shows that two components of the delay vector come closer but do not merge for $\dot{q}'$ time series. 
Figure \ref{fig:exp_multivariate}(c) shows that for the time series of $p'$, the two delay components adjust initially but stay separated. 
Hence the final dimension estimated for the regime of intermittency is $D_{final}=2+2=4$. 
For the dynamical regime of LCO, Fig. \ref{fig:exp_multivariate}(b) shows that two components of the delay vector associated with the time series of $\dot{q}'$ merges and similar behavior is observed in Fig. \ref{fig:exp_multivariate}(c). 
Thus the resultant dimension of the estimated phase space is $D_{final}=1+1=2$ for the regime of LCO. 

The comparison of delay values estimated using MAPSR and PECUZAL method for multivariate time series from turbulent combustor is shown in Table \ref{tab:exp_multi}. 
The delay values estimated for different dynamical regimes are shown separately for each of the time series of $\dot{q}'$ and $p'$. 
For the dynamical regime of chaos, the MAPSR method estimates the embedding dimension as 9, where 5 delay coordinates are from the time series of $\dot{q}'$ and 4 coordinates are from the time series of $p'$. 
On the other hand, for the regime of chaos, the PECUZAL  method estimates the embedding dimension of 5 as that of the univariate case with a single delay  coordinate from $\dot{q}'$ time series and 4 delay coordinates from time series of $p'$. 
For the regime of intermittency, the MAPSR estimates the four-dimensional delay embedding with 2 delay coordinates from $\dot{q}'$ and $p'$ time series individually. 
For the same dynamical regime, the PECUZAL method estimates six-dimensional delay embedding with no delay coordinate from the time series of $\dot{q}'$, and all the delay values estimated for $p'$ time series are the same as that of univariate case. 
For the regime of LCO, the MAPSR method estimates the delay embedding of dimension 2 with a single delay coordinate from each time series of $\dot{q}'$ and $p'$, whereas the PECUZAL method estimates $5$ dimensional phase space with no delay coordinate from $\dot{q}'$ time series and the delay values for $p'$ time series are same as that of univariate case.

The discrepancy in the estimated delay coordinates can be understood based on the idea behind the MAPSR and the PECUZAL method. 
The PECUZAL method is based on noise amplification, whereas the MAPSR method optimizes the delay embedding for modeling. 
In the case of multivariate time series data, the PECUZAL method might not include the time series causing noise amplification. 
This behavior of the PECUZAL method can be observed for the dynamical regimes of intermittency and LCO where there is no delay coordinate from the time series of $\dot{q}'$. 
On the other side, the MAPSR method optimizes the loss function that quantifies the prediction error.
Here, the loss function gives equal weightage to the prediction of all the delay coordinates. 
Hence, none of the time series is dropped out as the modeling is being performed to improve the prediction for input time series. 
Thus, one should apply the MAPSR method to those time series for which modeling is intended.  

The comparison of the true trajectories constructed using time series of $(\dot{q}', p')$ with the trajectories predicted using the MAPSR method with neural ODE model is shown in Fig. \ref{fig:exp_phase_multi}. 
For the dynamical regimes of chaos and intermittency, Fig. \ref{fig:exp_phase_multi}(a) and Fig. \ref{fig:exp_phase_multi}(b), the attractor is visualized in three-dimensional space with one delay component from the time series of $\dot{q}'$ and two other components from the time series of $p'$. 
Figure \ref{fig:exp_phase_multi}(c) shows the two-dimensional phase portrait for the dynamical regime of chaos, with one delay coordinate from the time series of $\dot{q}'$ and another from the time series of $p'$. 
We can see that the predicted trajectories are able to follow the true trajectories.

The comparisong of the MAPSR and PECUZAL method using the evolution of the normalized average deviation $\bar{\delta}_{cum}/\delta_{\max}$ for the time series of $\dot{q}'$ and $p'$ with normalized time $t/T_{\max}$ for the dynamical regimes of chaos and intermittency are respectively shown in Fig. \ref{fig:exp_loss_multi}(a) and Fig. \ref{fig:exp_loss_multi}(b).
The $\delta_{\max}$ is computed separately for each time series, and $T_{\lambda}$ is computed using time series of $p'$. 
The neural ODE with the optimal configuration obtained using the MAPSR method is used to predict for nearly $25T_\lambda$ duration for the dynamical regimes of chaos and intermittency. 
The deviation for the dynamical regime of chaos for MAPSR method initially grows slower compared to PECUZAL method for both $\dot{q}'$ and $p'$.
For the regime of intermittency, the deviation for the MAPSR method grows faster compared to PECUZAL method where PECUZAL method only considers time series of $p'$ from input time series of $\dot{q}'$ and $p'$.
Comparing the time scales over which the deviation grows for multivariate time series data (Fig. \ref{fig:exp_loss_multi})  with the univariate time series (Fig. \ref{fig:exp_loss_uni}), we can see that for the univariate case, the deviation stays lower for a longer duration compared to the multivariate case. 
This might be due to the noise amplification caused by the inclusion of time series of $\dot{q}'$ which has been discarded by the PECUZAL method for the dynamical regimes of intermittency and LCO (refer Table \ref{tab:exp_multi}). 

\subsection{Analysis of the observed results}\label{subsec:analysis}

The plot of the loss with initial dimension $d_{init}$, shows that loss initially decreases, attains minima at $d_{init}^{(opt)}$, and then increases. 
Similar behavior has also been observed by \citet{paper:young_2023}, where the first minima of AMI($\tau$) was used as a common difference $\Delta\tau_{AMI}$, and the dimensions for which prediction loss is minima was used to construct $UTDE$. 
The increase in the noise for clean data can be attributed to numerical error or to the discrepancy between the true and determined model. 
The increase in the loss for noisy time series after optimal dimension can be attributed to the high dimensionality of the noise. 
The reconstructed vector can be decomposed into a clean signal ($\vec{x}_{clean}$) and noise ($\vec{\epsilon}$), $\vec{x} = \vec{x}_{clean}+\vec{\epsilon}$. 

Increasing the dimension in steps resolves the trajectories, and the model can better approximate the trajectories. 
This might be the reason for the initial decrease in the loss by increasing $d_{init}$. 
At optimal dimension ($D_{opt}$), the trajectories $\vec{x}_{clean}$ are approximately captured by the model. 
Further, increasing the dimension does not convey significant information about trajectories. 
However, the added dimension leads to the addition of noise associated with an added component. 
The model will not capture this noise, and loss starts increasing with the addition of dimensions beyond the optimal dimension. 

We observe that, the training of the model with phase space which is of higher dimensions than $D_{opt}$, shows fewer modifications in the delay vector while training. 
Beyond $D_{opt}$, the trajectories are well resolved and are easier to capture by model and thus need fewer modifications in the delay coordinates. 

\section{Conclusion}\label{sec:conclusion}
The proposed MAPSR method combines phase space reconstruction with reduced-order modeling. 
It is a differentiable version of a time-delay embedding that can be jointly optimized with data-driven models such as neural ODEs. 
The minimization of the loss function with respect to model parameters and delay vector provides a model for the dynamical system as well as optimizes the delay embedding. 
The delay values can take values that are non-integer multiples of the sampling time as opposed to the existing methods that can take only integer multiples of the sampling time. 
For all the univariate cases that we tested, MAPSR has predicted the least embedding dimension except for the clean time series from the Lorenz system. 
From the Lorenz system, we can see that the MAPSR estimates the expected embedding dimension as 3 with the addition of noise. 
Though the addition of noise is not a requirement, the MAPSR method estimates smaller dimensions for time series with slight noise which is mostly the case with real-world time series data. 
With the application of the MAPSR to different dynamical regimes of the turbulent combustor, we demonstrated the generalizability of the method to real-world time series data. 
For multivariate time series from the same turbulent combuster, the MAPSR method predicted the same number of dimensions as that of univariate time series data, except for the regime of chaos. 
With the objective of modeling the dynamics which is equivalent to capturing the trajectories reconstructed using input time series, the MAPSR method optimizes the delay values for all input time series, whereas the PECUZAL method can drop some of the time series to reduce noise amplification. 
The neural ODE model trained using the MAPSR method is able to predict the true signal for nearly 7 to 8 Lyapunov time scales for the Lorenz system which is much better compared to the AMI-FNN and PECUZAL method for the same set of hyperparameters. 
For the univariate time series from the turbulent combustor, the average cumulative deviation initially grows faster for the MAPSR method but then stays in between PECUZAL and AMI-FNN methods. 
However, as for the dynamical regime of intermittency, MAPSR performs best.
For the multivariate time series from the turbulent combustor, the average cumulative deviation for the MAPSR method is lower than PECUZAL whereas the PECUZAL method estimates the delay embedding with a single time series and performs better than the MAPSR method for the intermittency regime. 

Here, we solely presented the combination of MAPSR with neural ODEs. 
However, it is a flexible approach that could be combined with other machine learning methods as well. 
The differentiability of MAPSR will result in an optimal phase space reconstruction in each of these cases. 
As such, we presented a strong building block for a data-driven approximation of dynamical systems. 
As we can inspect the learned PSR, this can also give further insights into the learned dynamics of the observed data.   

\begin{acknowledgments}
Jayesh Dhadphale is indebted to the Ministry of Human Resource Development for providing the fellowship under The Prime Minister's Research Fellows (PMRF) scheme and the International Immersion Experience Award, Office of Global Engagement, IIT Madras. 
R. I. Sujith acknowledges the funding from the IOE initiative (SB/2021/0845/AE/MHRD/002696), IIT Madras, India. 
Maximilian Gelbrecht acknowledges funding from the Volkswagen Foundation.
\end{acknowledgments}

\appendix
\section*{Appendixes}
\subsection{Algorithm of MAPSR method}\label{app:MAPSR_algo}
The Algorithm \ref{alg:MAPSR} describes the steps in MAPSR method.

\begin{algorithm*}[!ht]
    \SetKwFunction{isOddNumber}{isOddNumber}
    \SetKwInOut{KwIn}{Input}
    \SetKwInOut{KwOut}{Output}

    \KwIn{\begin{itemize}
          \item $\vec{S}=[s_i(t_j)]_{1\leq i\leq m, 1\leq j \leq N}$: Multivariate time series matrix with $m$ components and $N$ sampling instances.
          \item $\tau_{th}$: Threshold for merging delay values.
          \item $\Delta \tau_{init}$: Initial guess for common difference for delay.
          \item $d_{max}$: Maximum dimensions for individual time series.
          \item $f(\vec{x}, W)$: A mathematical model for the dynamical system (like neural ODE) which outputs $\dot{\vec{x}}$ and list of trainable parameters $W$.
          \item $iter_{max}$: Number of training iterations.
          \item $T_R$: Prediction horizon.
        \end{itemize}}
    \KwOut{$\vec{\tau}_{opt}$, $W_{opt}$}
    
    Initialize lists for saving outputs of training, $Loss_{train}$=[\ ], $\vec{\tau}_{train}$=[\ ], $W_{train}$=[\ ]
    
    \For{$d_{init} \leftarrow 1$ \KwTo $d_{max}$}{
        Initialize a delays for each of the $m$ time series with $d_{init}$ values, $\vec{\tau} = [\tau_{1,1},\cdots,\tau_{1,d_{init}},\cdots,\tau_{i,j},\cdots,\tau_{m,d_{init}}]$, where $\tau_{i,j}=(j-1)\Delta \tau_{init}$

        $\vec{X}$ = $interpolate$ ($\vec{S}$, $\vec{\tau}$) 

        Initialize $W$ with random values.

        \For{$iter \leftarrow 1$ \KwTo $iter_{max}$}{
            Choose a batch of trajectories of time duration $T_R$ as $\vec{X}_{batch}$.

            Compute the batch of predicted trajectories $\vec{X}_{pred}$ by time integrating $f$, with initial conditions for the trajectories same as $\vec{X}_{batch}$ and using $W$.

            Compute the loss function $\mathcal{L}$ using $\vec{X}_{batch}$ and $\vec{X}_{pred}$.

            Use automatic differentiation to compute $\nabla_{\vec{\tau}}\mathcal{L}$ and  $\nabla_{W}\mathcal{L}$.

            Use optimizer (eg. Adam, RMSprop)  to update $\vec{\tau}$, $W$.
            
            \For{$\tau_{i,j}$ in $\vec{\tau}$}{
                    $\vec{\tau}_{neb}$ = Set of all elements (excluding $\tau_{i,j}$) of $\vec{\tau}$ corresponding to $i^{th}$ time series whose seperation from $\tau_{i,j}$ is less than $\tau_{th}$.

                    Remove $\vec{\tau}_{neb}$ from the $\vec{\tau}$ \tcp*[f]{Delay merging.}

                    Modify $W$ to accommodate the change in $\vec{\tau}$.
            }
        }   

        Save training results for $d_{init}$ as ${Loss}_{train}[d_{init}]=\mathcal{L}$, $\vec{\tau}_{train}[d_{init}]=\vec{\tau}$, $W_{train}[d_{init}]=W$
    }

    Find $d_{init}$ for which ${Loss}_{train}$ is minimum as $d_{init,opt}$

    Assign, $\vec{\tau}_{opt}=\vec{\tau}_{train}[d_{init,opt}]$ and $W_{opt}=W_{train}[d_{init,opt}]$

    \KwRet{$\vec{\tau}_{opt}$, $W_{opt}$}
    \caption{MAPSR algorithm}
    \label{alg:MAPSR}
\end{algorithm*}

\section*{Data Availability Statement}

Data will be made available on reasonable request to the corresponding authors. Code for MAPSR method is publicly available at \url{https://github.com/JayeshMD/MAPSR.git}

\bibliography{aipsamp}
\end{document}